\documentclass[letterpaper, 11pt]{article}
\usepackage[utf8]{inputenc}

\usepackage{booktabs} 
\usepackage{amsmath,amssymb,amsthm,algpseudocode,algorithm,float,tkz-berge,caption,mathtools}
\usepackage{graphicx, float}
\usepackage[colorinlistoftodos]{todonotes}
\usepackage[margin=1in]{geometry}
\usepackage{balance}

\newcommand{\E}{\mathbb{E}}
\newcommand{\p}{\mathbb{P}}
\newcommand{\OO}{\mathcal{O}}

\newtheorem{theorem}{Theorem}
\newtheorem*{theorem*}{Theorem}
\newtheorem{lemma}{Lemma}
\newtheorem{definition}{Definition}
\newtheorem*{definition*}{Definition}
\newtheorem*{lemma*}{Lemma}
\newtheorem{corollary}{Corollary}
\newtheorem*{corollary*}{Corollary}

\newtheorem*{claim*}{Claim}
\newtheorem{proposition}{Proposition}
\newtheorem*{proposition*}{Proposition}

\begin{document}
	\title{The Cost of Uncertainty in Curing Epidemics}
	
	\author{Jessica Hoffmann \and Constantine Caramanis}
	\maketitle
		
	\begin{abstract}
		Motivated by the study of controlling (curing) epidemics, we consider the spread of an SI process on a known graph, where we have a limited budget to use to transition infected nodes back to the susceptible state (i.e., to cure nodes). Recent work has demonstrated that under perfect and instantaneous information (which nodes are/are not infected), the budget required for curing a graph precisely depends on a combinatorial property called the {\sc CutWidth}. We show that this assumption is in fact necessary: even a minor degradation of perfect information, e.g., a diagnostic test that is 99\% accurate, drastically alters the landscape. Infections that could previously be cured in sublinear time now may require exponential time, or orderwise larger budget to cure. The crux of the issue comes down to a tension not present in the full information case: if a node is suspected (but not certain) to be infected, do we risk wasting our budget to try to cure an uninfected node, or increase our certainty by longer observation, at the risk that the infection spreads further? Our results present fundamental, algorithm-independent bounds that tradeoff budget required vs. uncertainty. 
	\end{abstract}
	
	%
	%


	\section{Introduction}
\label{sec:intro}

Epidemic models are used across biological and social sciences, engineering and computer science, and have had important impact in the study of the dynamics of human disease, computer viruses, but also trends rumors, viral videos, and most recently the spread of fake news of social networks. Their history in the literature dates to the first mathematical model of epidemics by Bernoulli in 1760 ~\cite{Bernoulli2004}.
In this paper, we focus on epidemics propagating on a graph, as introduced by the seminal paper ~\cite{Newman2002}. In particular, we consider so-called SI models (see below for a precise definition) where an infected node can only propagate the infection to its non-infected neighbor, as opposed to the fully mixed models considered in the early literature. This graph-based approach provides a more realistic model, in which the spread of the epidemic is determined by the connectivity of the graph, and accordingly some nodes may play a larger role than others in the spread of the infection.  

At any point in time, the {\em state} of an SI-type epidemic on a graph is given by the list of nodes on the graph that are infected, and their relative topology (position) in the graph. Having a good estimate of the state is critical, as it determines the dynamics of the spread of the epidemic into the future. As a simple example, we can ask what the spreading rate is on an $N$-node line graph of an infection with $N/2$ infected nodes. If those nodes are contiguous, then it will take $O(N)$ time for the epidemic to spread to the entire graph. If every other node is infected, it will take $O(1)$ time. 

If we have access to the status of each node (infected or not), then we know the state exactly. Much work has focused on the state estimation problem, in the setting where only noisy information is available. Indeed, work in ~\cite{Milling2015,Milling2015a,meirom2015}, ~\cite{Arias-castro2011,Arias-Castro2008,Arias-castro}, and elsewhere, considers a setting where only noisy observations of the status of each node are possible, and even answering whether there is an epidemic or not is a challenge. Those and related works, as we discuss in more detail below, focus on the problems related to epidemic state estimation, and do not consider the control problem of curing the epidemic.

%

On the other side, the problem of curing an epidemic with a limited budget, {\em but with perfect observation} (i.e., perfect knowledge of the state at each point in time), has been recently considered in ~\cite{Drakopoulos2015a,Drakopoulos2014}. Their budget, as we explain more precisely further below, is essentially a bound on the {\em curing effort} they can expend at a given time (as opposed to total curing effort over time). In this setting, the problem is to optimize the allocation of the curing budget across nodes at every point in time. They characterize the budget required for fast curing, as a function of a combinatorial property of the graph -- its {\sc CutWidth} (we define this below).

The problem of curing an epidemic with a limited budget and partial observation of the state of the epidemic (i.e., which nodes are infected and which are not) introduces a fundamentally new element to the problem. Indeed, this interaction represents a fundamental tension: our estimate of the state of a node improves the longer we observe it, and so the longer we wait to cure a node, the less likely we are to waste precious curing resources on non-infected nodes. On the other hand, the longer an infected node remains untreated, the more the epidemic spreads. To the best of our knowledge, no work has successfully attacked the problem of curing an epidemic with a limited budget and partial observation of the state of the epidemic (i.e., which nodes are infected and which are not). Our work considers precisely this problem, and therefore, broadly speaking, is about the interaction of -- specifically, simultaneous -- learning and control. 

By considering learning the state and controlling the epidemic simultaneously, we prove a lower bound that shows (see Section \ref{sec:partial} for precise result) that partial information can have a dramatic impact on the resources (either time or budget) required to cure an infection: even with slightly imperfect/incomplete information, the time to cure a particular graph may increase exponentially, unless the budget is also significantly increased. Concretely, we show that if instead of receiving the state of each node at each point in time, we receive a slightly noisy (e.g., only 99\% accurate) guess of the state, then there is no constant factor of the {\sc CutWidth} which is sufficient for {\em any algorithm} to cure the epidemic in linear (expected) time. 
 
\subsection{Related Work and Background}
\label{sec:related}


Detecting an epidemic, as well as its location, under noisy data, has been well-studied in ~\cite{Arias-castro2011}, in the context of detecting a multidimensional anomalous cluster, with time playing the same role as any other dimension. Graph-specific epidemic detection has been further studied by ~\cite{Sharpnack2012}, with constraints based on the cut of this anomalous cluster. ~\cite{Milling2015} study the detection of epidemic-specific clusters by detecting the shapes which arise specifically when there is an epidemic. The focus in those works has been to understand the limit of information required in order to detect the epidemic. More generally, inverse problems have also been of interest, especially source detection ~\cite{spencer2015impossibility,shah2011rumors, shah2012rumor,wang2014rumor, shah2010detecting} or obfuscation ~\cite{fanti2015spy, fanti2016rumor}. 

In our work, we adopt a much stronger observation model than in the papers listed above; our negative result establishes, however, that controlling the epidemic is impossible with weaker information than the threshold we characterize.

In ~\cite{Drakopoulos2015,Drakopoulos2014}, the authors tackle the problem of curing graphs with perfect knowledge of the state of each node, constrained by a budget which corresponds to the speed at which the nodes are cured. Their results show that there exists a threshold phenomenon: for any given graph, if the curing budget is lower than a combinatorial quantity of the graph called the {\sc CutWidth}, the curing time is exponential; if it is higher, they exhibit a strategy to cure any graph in sublinear time. The {\sc CutWidth} captures a key bottleneck in curing, and is important in our work as well. Therefore it is useful to define this precisely now.
\begin{definition}
\label{def:cutwidth}
Given a graph $G = G(V,E)$, and any subset of the nodes, $S \subseteq V$, the {\sc Cut} of $S$ is the number of edges crossing from $S$ to $S^c$. Given any sequence of $|V|+1$ subsets $S_0,\dots,S_{|V|}$ such that $S_0 = \emptyset$, $S_{|V|} = V$, and $S_k$ and $S_{k+1}$ differ by the addition of a single node (called a {\em crusade} in \cite{Drakopoulos2014,Drakopoulos2015}), the cut of the sequence is the largest cut of any of the sets $S_k$. The {\sc CutWidth} of a graph is the minimum cut of any sequence satisfying the above properties.
\end{definition}
Intuitively, the {\sc CutWidth} of a graph is the largest cut one would be {\em forced} to encounter when curing a graph. The cut of a subset is critical, because for an infected set of nodes $S$, its cut is the number of non-infected nodes adjacent to infected nodes, and hence is the instantaneous rate of infection of the epidemic at that moment (in that configuration). For an illustration, consider again an $N$-node line graph. Its {\sc CutWidth} is equal to one, since when curing the graph from one end to the other, we have only one single non-infected node adjacent to an infected node at any time. Note that this is the best case, because if we were to start curing nodes in the middle of the infection, the cut between the infected nodes and the non-infected nodes could be made as large as $O(N)$. 

Their strategy is based on two main ideas. The nodes are cured following an ordering which keeps the cut between the infected set and the non-infected set as low as possible. Then, as soon as there is a new infection, the strategy switches to damage control, and focuses on returning to the ordering previously mentioned. 

Our result hinges on the fact that the damage-control part of the strategy is exactly the part which is hard to accomplish with partial information. If the number of $k$-hop neighbors of a node grows exponentially, as is the case for the binary tree, detecting where the infection can have spread becomes a difficult task. Moreover, if we can detect such an escape path, but the infection has spread to a high number of nodes by the time we have enough information to try to prevent it, detection was useless. It is the tension between waiting less time and wasting budget on false alerts, or waiting too long and being unable to prevent the spread, which makes the problem of curing with partial information challenging.

	\section{Model and main contributions}
The key elements that define our model are the dynamics of the spreading process and the controlled curing process, and then the stochastic process that defines the degradation from perfect information. We describe these in detail, in this order. We then provide a few basic definitions that appear repeatedly throughout the paper, and then finally outline the main contributions of this work. 

\subsection{The SI + curing model}
\label{model}
In a standard SI (susceptible $\rightarrow$ infected) model, an epidemic spreads along edges from infected nodes to their neighbors according to an exponential spreading model: when a node becomes infected, it infects each uninfected neighbor according to an exponential random variable. SIS models are SI models where infected nodes also transition to susceptible, again at an exponential rate. Here, we consider the setting where the rate at which nodes transition from infected to susceptible is under our control, subject to a budget. How to optimally use this budget is the main question at hand. We prefer to call this a {\em controlled SI} process rather than a SIS process, because we are interested in the regime where our total curing budget is $o(N)$, where $N$ is the number of nodes. A SIS process typically has transitions from susceptible to infected of the same order as the infection rates; in our setting, this would correspond to a budget of at least $O(N)$. We note that much work has considered this setting, and has characterized the absorption time (into the ``all cured'' state) as a function of the topology of the network \cite{ganesh2005effect}.

In the sequel, we consider a discrete, Bernoulli approximation to these exponential rate models, by considering the dynamics evolving with discrete time steps $\tau$; {\em we then take the time step $\tau$ to zero}, hence recovering the continuous time dynamics. In particular, this model is a discretization of the exponential model of \cite{Drakopoulos2015a}. As $\tau \to 0$, the models become equivalent. This discretization and the subsequent limit as $\tau \rightarrow 0$ facilitate our quantification of uncertainty, i.e., how much information we receive about the state of each node, in a given time interval. This is defined precisely below.

The dynamics of this controlled stochastic process evolve as follows. At each time $t$, for all $N$ nodes of the graph, the decision-maker assigns a budget $r_i^t$, subject to the constraints $\sum_{i =1}^{N} r_i^t = r$. During a time step of length $\tau$, each node $i$ is cured with probability $\delta_i^t = 1 - e^{-r_i^t \tau}$ if it was infected, and nothing happens otherwise -- the budget is wasted. Then, for every edge between an infected and a susceptible node, an infection occurs with probability $\mu = 1 - e^{-\tau}$. The number of infected nodes at time $t$ is given by $I_t$. In particular, since the graph is completely infected at the beginning, we have $I_0 = N$. We summarize the notation in Table \ref{notation-table}.  

\begin{figure}
	\centering\includegraphics[width=7cm]{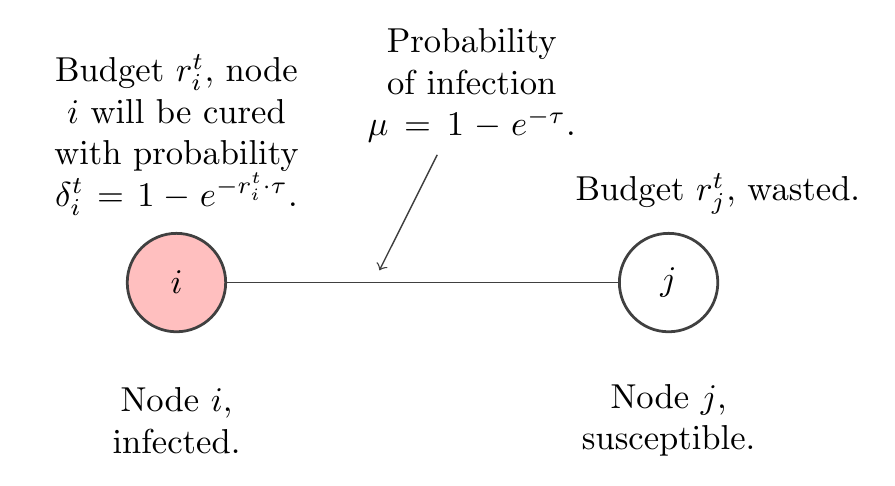}
	\caption{Visual representation of the different parameters -- see Table 1 for more details}
\end{figure}

\begin{table*}
  \caption{Notations}
  \label{notation-table}
  \begin{tabular}{cl}
    \toprule
$N$  & Number of nodes in the graph \\
$I_t$ & Number of infected nodes at time $t$ \\
$\tau$ & Size of a time step \\
$r_i^t$  & Budget spent on node $i$ at time $t$\\
$r = \sum_{i = 1}^{N} r_i^t$  & Total budget for each time step \\
$\mu =  1 - e^{- \tau}$ & Probability of an infection along an edge between a susceptible and an infected node \\
$\delta_{i}^{t} = 1 - e^{-r_{i}^{t} \cdot \tau}$ & Probability that node $i$ gets cured at time $t$ (if already infected)\\
$\delta = 1 - e^{-r \cdot \tau}$ & Maximum probability of being cured for a node \\
$p$ & $\mathbb{P}(\text{node $i$ raises a flag at time t } | \text{ node $i$ is infected})$ \\
 $q$ & $\mathbb{P}(\text{node $i$ raises a flag at time t } | \text{ node $i$ is susceptible})$  \\
    \bottomrule
  \end{tabular}
\end{table*}


We now give a few definitions related to the above quantities, that we use throughout this paper.

\begin{definition}
We call \textbf{curing process} the stochastic process of cures and infections according to the model described in section \ref{model}. This process has a deterministic part (how much of the budget is assigned to which nodes at each time step), and a stochastic part (curing and infection follow geometric laws).
\end{definition}

\begin{definition}
We call a \textbf{strategy} the set of budgets assigned for each node at each time: $\{ r_i^t,\, i\in [N],\, t = k\cdot \tau, \, k \in \mathbb{N}\}$. We note that in the Partial Information setting that we introduce below, the actions taken at time $t_1$ may depend on the information accumulated until time $t_1 - \tau$. 
\end{definition}

In the rest of the paper, we refer to the set of infected nodes (resp. susceptible nodes) as the \textbf{infected set} (resp. \textbf{susceptible set}). We may also refer to the cut between the infected set and the susceptible set as the \textbf{cut}. When we use the word \textbf{distance} between two nodes in a graph, we refer to the number of nodes in the shortest path between these two nodes. The distance between a node and a set is the shortest distance between this node and any node of the set.

\subsection{Partial Information/Blind Curing}
In the Complete Information setting, we assume that the status (infected or susceptible) of each node is known at each point in time. In what we call the Blind Curing model, we never have any information about the status of each node. The Blind Curing model is a technical tool we use en route to the final result. We introduce a Partial Information model that interpolates between these two extremes, and indeed is our main object of interest. Our model of partial information provides a stark tradeoff for the decision-maker: allocate resources to nodes whose status is very uncertain, and thus significantly raise the probability of wasting curing resources, or wait to collect more information and hence more certainty about the status of a node, running the risk that an infected node was allowed to infect neighbors unfettered. 

Our motivation for our partial information model comes from zero-day {\em behavioral} malware detectors, often called {\em Local Detectors} \cite{bose2008,gregoire2008}, where anti-malware software raises alerts of ``suspicious behavior'' that are then related to a central authority. We refer to these alerts as ``flags.'' Thus, in the Complete Information model, an infected node would raise a flag at each instant with probability $1$, and an uninfected node would never raise a flag. In the \textbf{Partial Information} model, at each time step, each node, independently of all others, raises a flag with some probability. The probability of getting a flag is $p$ if the node is infected, $q$ if the node is susceptible, with $p>q$. By aggregating the information about a node over multiple time steps, we can use basic concentration inequalities to deduce its state, and thus more observation time corresponds to higher certainty about a node's state.

As noted above, $p=1, q=0$ recovers the \textbf{Complete Information} setting, and $p=q$ the \textbf{Blind Curing} setting. 

In order to recover the continuous time dynamics, we let $\tau \rightarrow 0$. The key quantity that measures the amount of information per fixed unit time, is given by the rate function from Sanov's theorem, normalized by the time step: $\frac{\mathcal{D}(p||q)}{\tau}$, where $\mathcal{D}(p||q)$ is the Kullback-Leibler distance between $p$ and $q$ \cite{cover2012elements}. To understand this intuitively, this says that when $\frac{\mathcal{D}(p||q)}{\tau}$ is a constant, observing a node for a fixed period of time corresponds to administering a test with a nonzero false positive and false negative probability. That is, we can know the state of a node with constant probability of error by observing this node over a constant amount of time, which is what one expects from a real-world source of information. Note that as $\tau \rightarrow 0$, if $p-q$ is constant (or, more generally, if $D(p||q)$ goes to zero sublinearly) then we recover the Complete Information setting. Hence, the setting of interest is where $(p - q) \rightarrow 0$ as $\tau \rightarrow 0$, and the critical scaling is controlled by $\mathcal{D}(p||q)/\tau$.

\subsection{Main contributions}
Our main result consists of two parts. First, we show that there exist graphs that cannot be cured in polynomial time in the Blind Curing model. We then use this result to get a lower bound for the cost of lack of information in the Partial Information model. We obtain an expression for the lower bound that shows the required tradeoff between $\frac{\mathcal{D}(p||q)}{\tau}$ (the information available per unit of time), and the budget, $r$.

\begin{theorem}{A Partial Information impossibility result.} \label{th:partialInfo} \\
	We consider the task of curing a fully infected complete balanced binary tree with $N$ nodes. Let $\frac{\mathcal{D}(p||q)}{\tau}$ be a measure of the amount of information we get per time step, and $r$ be the budget (curing rate) of our curing process. If 
\begin{equation}
\frac{\mathcal{D}(p||q)}{\tau} \leq  \OO\left(\frac{\log(N)\sqrt{\log(r)}}{r }\right),\label{eq:mainresult}
\end{equation}
as $\tau \to 0$, then it is fundamentally impossible for any algorithm (of any computational complexity) to cure the complete binary tree in polynomial expected time with budget $r = \OO(W^{\alpha})$, where $W$ is the {\sc CutWidth} of the graph and $\alpha$ is any constant.
\end{theorem}

For the Blind Curing case, we also have the following {\em upper bound}.
\begin{theorem}\label{thm:new_upperbound} For all $c> 0$, we can always cure the binary tree in expected linear time with budget $\OO(e^{4/c}N^c)$. In particular, our strategy does not require any information about the state of the nodes.
\end{theorem}

{\bf Interpreting the result}. Suppose that if a node is observed for a fixed period of time, we can estimate its state (infected or not) with probability $1-\delta$ for $\delta$ some constant. Our results say that regardless of what this constant is, e.g., even if we have a test that takes 1 minute (or other time unit) to implement and returns a result that is $99\%$ (or any other constant quantity) accurate, then polynomial time curing is impossible, for budget any multiple of the {\sc CutWidth}. Indeed, as explained above, a constant-error estimate in a fixed unit of time corresponds to $D(p||q)/\tau$, the left-hand side of (\ref{eq:mainresult}), being a constant. On the other hand, if the budget is any multiple of the {\sc CutWidth}, the right-hand side of (\ref{eq:mainresult}) grows like $\sqrt{\log\log(N)}$, and in particular is larger than any constant. In contrast, with complete (and instantaneous) certainty of the state of each node (here the left-hand side of (\ref{eq:mainresult}) can be infinite), ~\cite{Drakopoulos2014} proves that every graph can be cured in linear expected time with budget higher than the {\sc CutWidth}.

For the blind setting, Theorem \ref{th:partialInfo} says that for budget of any polynomial of $\log(N)$, curing takes superpolynomial time. Theorem \ref{thm:new_upperbound} gives an upper bound that shows that this lower bound is not too far off; it says that a budget of $N^c$ is sufficient, for any $c > 0$. This theorem is proved in Appendix \ref{sec:upperBound}.

Our result focuses on the binary tree. Since our main result is a {\em lower bound}, this specific example is sufficient to resolve the question of whether the {\sc CutWidth} (or something proportional to it) is the right quantity to focus on to build a curing strategy robust to noise in our node estimates. In addition to this, we note that many graphs contain trees as subgraphs. Since adding nodes and edges only makes curing more difficult, our results can be seen to apply to any graph structure with a binary tree as a subgraph (as long as adding edges does not dramatically change the {\sc CutWidth} of the graph).
%

{\bf Proof Idea}. Our proof focuses on bottlenecks of the curing process: events that {\em must happen with high probability, regardless of the policy used, en route to curing an infection.} Specifically, our proof hinges on showing two such bottlenecks. First, we show that regardless of the policy, regardless of the stochastics of the curing and infection process, with high probability the last nodes to be cured cannot all be far from the root node. As we discuss below, the intuitive reason for this relies on our graph topology, and the fact that the cut between the set of infected nodes and the set of uninfected nodes must remain low if we hope to control the infection. On a binary tree, a simple calculation (Proposition \ref{cl:rootclose}) shows that any $\frac{N}{r^4}$-node set with low cut must contain nodes close to the root. The significance of this result is that at all times that matter (namely, at all points where the curing policy might be close to succeeding), there will be infected nodes that are not far from (exponentially) many uninfected nodes. Next, we show that in any interval of time, there must be many uninfected nodes {\em that are also unprotected} by the curing policy, regardless of what the curing policy is doing (Lemma \ref{lem:minimalTree}). In Theorem \ref{th:BlindCuring}, we combine these results to show that the probability that an infection begins, travels through the root to the unprotected subset of nodes and infects them before the remaining nodes are cured, is very close to 1.

	\section{Proof sketch}

\begin{figure*}
\centering\includegraphics[width=16cm]{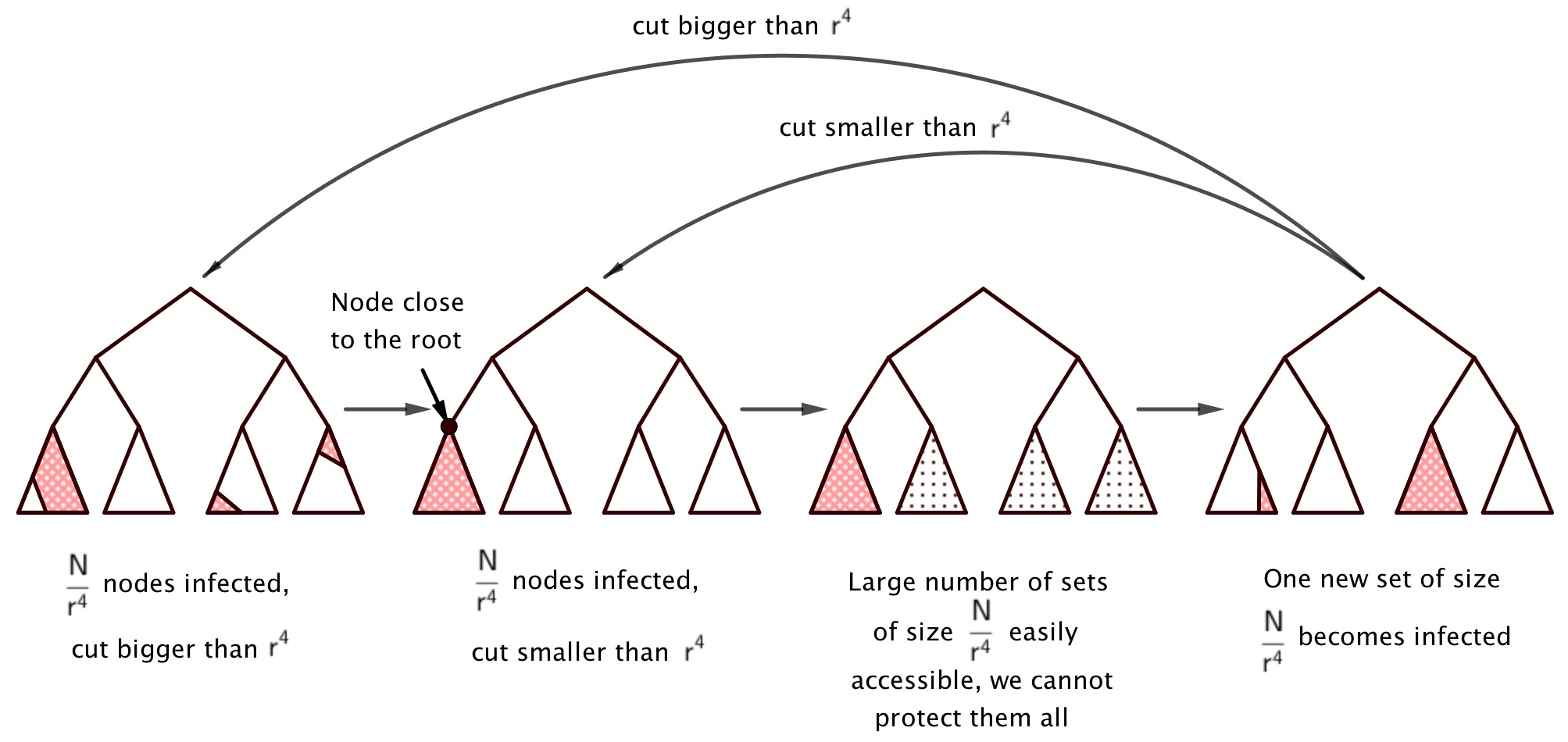}
\caption{Visual representation of the main steps of the proof: when only $\frac{N}{r^4}$ nodes remain infected, no strategy can prevent the reinfection of $\frac{N}{r^4}$ new nodes in some other part of the graph. The graph can only be cured if the cycle is broken, a rare event which takes superpolynomial time in expectation.}
\end{figure*}
We first prove that polynomial curing is impossible in the Blind Curing setting if the budget is polynomial in the {\sc CutWidth}. We then show that in the Partial Information setting, we do not obtain enough information to detect threats of reinfection, and thus cannot prevent them: we are "blind" to the threats until it is too late.

Our proof in the Blind Curing setting focuses on a subprocess which is bound to happen for any curing strategy. We consider the last $\frac{N}{r^4}$ infected nodes. We show that by the time we cure these last remaining infected nodes, a new set of $\frac{N}{r^4}$ nodes becomes infected with high probability. Trying to cure the whole graph is then similar to playing a very long game of whack-a-mole with superpolynomial expected end time.

\subsection{Blind Curing setting}
\begin{description}
\item[Step 1 (Section \ref{sec:step1}):] We first show that if a strategy allows the cut between the infected and susceptible set to be much higher than the available budget $r$, the infection becomes uncontrollable. In this case, the infection rate exceeds the curing rate, and the reinfection would be inevitable even if we had complete knowledge about the infection state of each node at each time (\textit{i.e.} this happens even in the Complete Information setting). In particular, if $\frac{N}{r^4}$ nodes are infected and the cut is above $r^4$, the drift of the curing process is dominated by the infections. We can then use random walks results, such as Wald's Inequality, to prove that after a few time steps, we end up with at least as many infected nodes, but a cut below $r^3$ (we actually end up with many more infected nodes, but as many is enough for the proof). We can therefore focus on analyzing the situation in which $\frac{N}{r^4}$ nodes remain infected with a cut lower than $r^4$.

\item[Step 2 (Section \ref{sec:step2}):] Due to the topology of the binary tree, a cut below $r^4$ implies that there exists an infected node which is close to the root. This makes it easy for the infection to escape through the root, and reach a large number of susceptible nodes. One key point of the proof is that this node will remain infected (and therefore potentially infecting) for a very long time, and an infection can start at any time step during this period.

\item[Step 3 (Section \ref{sec:BlindCuring}):] Since the infection escapes through the root, the number of uninfected nodes easily accessible is very large, and specifically, larger than the budget. This makes it impossible to cover all the potential escape routes. Notice that this is very specific to the Blind Curing setting: if we knew in which direction the infection was escaping, we could prevent it as in ~\cite{Drakopoulos2015a}. It is because the number of potential infected nodes is exponentially higher than the number of actual nodes infected, and because we do not know where the infection actually is, that we end up wasting considerable curing budget on uninfected nodes. Therefore, the infection is very likely to escape, and a new set of $\frac{N}{r^4}$ nodes becomes infected again.
\end{description}

\subsection{Partial Information setting}
To extend this result to the Partial Information setting, we notice that as soon as the cut of the new infection reaches $3r$, we can use Gambler's Ruin results to show that at least $\frac{N}{r^4}$ nodes will become infected with constant probability. If we cannot detect the infection escaping until a cut of $3r$ is reached, we therefore cannot prevent the reinfection with constant probability. Using Sanov's Theorem, we show that the uncertainty in our state estimation for any node does not resolve itself quickly enough (in particular, with respect to how fast the neighborhoods of the binary tree grow). Specifically, the infection remains undetectable with constant probability until a cut of $3r$ is attained. This allows us to extend the result from the Blind Curing Setting to the Partial Information setting.

	\section{Proof for Blind Curing} 
In this section, we prove that in the Blind Curing setting, we cannot cure a complete binary tree in polynomial time with budget $\OO(W^\alpha)$, where W is the {\sc CutWidth}, and $\alpha$ is any constant. A complete binary tree has {\sc CutWidth} smaller than $\log(N)$ (Proposition \ref{cl:Wtree} in the appendix). Therefore, in the rest of the paper, we set $r = \log^\alpha(N)$.

We focus on the last moments of the curing, when only $\frac{N}{r^4}$ nodes remain infected. The proof relies on the fact that by the time we cure these last $\frac{N}{r^4}$ nodes, a new set of $\frac{N}{r^4}$  nodes will have become infected in another part of the graph with probability superpolynomially close to 1.

\subsection{The infection cannot be controlled when the cut is too high} \label{sec:step1}
We start by proving that without loss of generality, we can suppose the cut between the infected set and the susceptible set is less than $r^4$ when $\frac{N}{r^4}$  nodes remain infected. If the cut is above $r^4$, the infection becomes uncontrollable with high probability, and we end up with at least as many nodes infected \footnote{It is actually more likely that a large number of new nodes will become infected. However, our proof only requires the total number of infected nodes to not decrease, so this is what we prove.}, but a cut below $r^3$ after some time steps. Therefore, supposing the cut is below $r^4$ only reduces the expected time of curing. 

Intuitively, if the budget is much smaller than the cut, the leading term in the drift of the infection process will be driven by the new infections taking place, regardless of the policy in use. Trying to eradicate, or even contain, an epidemic in these conditions would be like fighting an avalanche with a flamethrower: some snow will melt, but it will not stop the avalanche - which will only stop by itself. Similarly, we can only hope to regain some control over the infection process when the budget is at least of the same order of magnitude as the cut. 

To prove this result, we introduce a random walk $G_t$ which stochastically dominates the curing process (Lemma \ref{lem:RW}). We define a stopping time, $T_{\rm SmallCut}$, which corresponds to the first time the cut reaches $r^3$.  We prove that by the time we reach this stopping point, many infections must have taken place (Lemma \ref{lem:Wald} and \ref{cl:GmanyInfection}), which implies that many time steps must have gone by. We can then use concentration inequalities to prove there are at least as many infected nodes at $T_{\rm SmallCut}$ as there were at the beginning of the random walk (Lemma \ref{lem:MoreCures}).

\begin{definition}
Let $A_t \sim \mathcal{B}(r,\delta)$, a binomial random variable with $r$ trials and probability $\delta$, and $B_t \sim \mathcal{B}(\frac{r^3}{3},\mu)$, a binomial with $\frac{r^3}{3}$ trials and probability $\mu$. 

We define the random walk $G_t$:
\[ G_t = \sum_{t'=t_0}^t  A_{t'} - B_{t'}.\]
\end{definition}

We are especially interested in the sign of the random variable 
$$G_{T_{\rm Small Cut}} =\displaystyle \sum_{t'=t_0}^{T_{\rm Small Cut}} A_{t'} - B_{t'}.$$ 
\begin{definition}
We call the \textbf{increase in susceptible} (uninfected) \textbf{nodes} since $t_0$ the random variable $I_{t_0} - I_t$, for $t > t_0$. This is the difference between the total number of infected nodes at time $t_0$, and the total number of infected nodes at time $t > t_0$. In other words, it corresponds to the difference between the number of nodes we successfully cured and the number of newly infected nodes between the times $t_0$ and $t$. Note that if more infections than curings have happened since $t_0$, the increase in susceptible nodes is negative.
\end{definition}

\begin{definition}
A random variable $X_1$ is \textbf{stochastically dominated} by a random variable $X_2$, if $\p[X_1 \geq x] \leq \p[X_2 \geq x]$ for all $x$. 
\end{definition}

\begin{lemma} \label{lem:RW} 
Let $t_0$ be the first time such that $I_{t_0} = \frac{N}{r^4}$ and the cut is above $r^4$. The random walk $G_t$, defined above, stochastically dominates the quantity $I_{t_0} - I_t$ (the increase in susceptible nodes since $t_0$) for any $t \leq T_{\rm Small Cut}$,  for every strategy.
\proof At each time step $t$, each node $i$ is assigned a budget $r_i^t$, with $r_i^t \leq r$, and gets cured with probability $\delta_i = 1- e^{-r_i^t \cdot \tau} \leq \delta = 1 - e^{-r \cdot \tau} $. By assumption of our model, there are at most $r$ nodes being cured, among which at most $r$ are infected (we do not know for sure if the nodes we are curing are infected or not since we are in the Blind Curing setting). Each of these infected nodes can therefore return to the susceptible state with probability at best $\delta$. In other words, the number of cured nodes is stochastically dominated by a binomial variable with $r$ trials and probability $\delta$, \textit{i.e.,} it is stochastically dominated by $A_t$.

Before the stopping time, the cut is at least as big as $r^3$. The maximum degree in a tree is 3, so 3 of these edges could lead to the same node. Therefore, there are at least $\frac{r^3}{3}$ potential infections happening with probability $\mu$. $B_t$ is therefore stochastically dominated by the number of new infections in the curing process, for any strategy. 

Thus, $G_t$ stochastically dominates $I_{t_0} - I_t$, for any $t \leq T_{\rm Small Cut}$, for every strategy.
\qed 
\end{lemma}

We use random walks properties to exponentially bound the probability that $G_{T_{\rm Small Cut}}$ is positive, which correponds to more cures than infections. We recall Wald's Inequality for random walks, whose proof appears in Section 9.4 of \cite{Gallager2013}.

\begin{theorem} {Wald's identity for 2 thresholds}  \label{th:wald2} \\
Let $X_i$, $i \geq 1$ be i.i.d. and let $\gamma(r) = \log(\E[e^{rX}])$ be the Moment Generating Function (MGF) of $X_1$. Let ${\rm Int}(X)$ be the interval of $r$ over which $\gamma(r)$ exists. For each $n \geq 1$, let $S_n = X_1 + \dots +X_n$. Let $\epsilon > 0$ and $\beta < 0$ be arbitrary, and let $J$ be the smallest $n$ for which either $S_n \geq \epsilon$ or $S_n \leq \beta$. Then for each $r \in {\rm Int}(X)$:
\[E[\exp(rS_J -J{\gamma}(r))] = 1. \]
\end{theorem}

\begin{corollary} \label{cor:wald} Under the conditions of Theorem \ref{th:wald2}, assume that $\E[X] < 0 $ and that $r^* > 0$ exists such that $\gamma(r^*) = 0$. Then:
 \[\p[S_J \geq \epsilon] \leq \exp(-r^*\epsilon). \]
\end{corollary}

We now use Wald's Inequality to prove $I_{t_0} - I_t$ cannot be very large.

\begin{lemma} \label{lem:Wald} 
If the cut is above $r^3$, the probability that the increase in susceptible nodes $I_{t_0} - I_t$ is higher than $K$ is exponentially small in $K$.
\proof
The curing process is stochastically dominated by the random walk described above.  Let $P_{\rm curingK}$ be the probability that $G_t$ reaches the value $K$ before stopping. 
Using Wald's Inequality (Corollary \ref{cor:wald}): 
$$ P_{\rm curingK} \leq e^{-x^*\cdot K}. $$
where $x^*$ is a value for which the MGF of $G_t$ is 1. We prove in Proposition \ref{cl:MGF} in the appendix that there exists such a $x^* > 0$, and in Proposition \ref{cl:MGFtau0}  in the appendix that $x^*$ converges to $\log(\frac{r}{3})$ when $\tau \to 0$.   \qed 
\end{lemma}

\begin{corollary} \label{cor:polylog}
The increase in susceptible nodes  $I_{t_0} - I_t$ is bounded above by $\frac{\log^2(N)}{x^*}$ with probability at least $1 - e^{-\log^2(N)}$.
\proof
Using Lemma \ref{lem:Wald}, we have:
\begin{align*}
e^{-x^*\cdot K} &\geq  e^{-\log^2(N)} \implies K \leq \frac{\log^2(N)}{x^*}.
\end{align*}
We conclude with setting $K = I_{t_0} - I_t$.
\qed 
\end{corollary}

We deduce from the previous result that many infections must have taken place.

\begin{proposition} \label{cl:GmanyInfection} 
At $T_{\rm SmallCut}$ (when the cut reaches $r^3$), at least $\frac{ r^4 }{7}$ infections will have taken place.
\proof Let C be the number of nodes cured between $t_0$ and $T_{\rm Small Cut}$ , and I be the number of new infections in the same time period. Any curing or infection reduces the cut by at most 3, since the graph is a binary tree. Therefore:
$$ 3C + 3I \geq r^4 - r^3.$$
On the other hand, using Corollary \ref{cor:polylog}, we can bound the increase in susceptible nodes:
$$ C - I \leq \frac{\log^2(N)}{x^*} .$$
Combining the two inequalities:
\begin{align*}
 I &\geq \frac{ r^4 - r^3}{6} - \frac{\log^2(N)}{x^*} \geq_{N \gg 1} \frac{ r^4}{7}.
\end{align*} 
\qed 
\end{proposition}

The previous Proposition proved that many infections happened. We now show this implies that many time steps must have passed by, which allows us to use concentration inequalities. To prove the next Lemma, we recall Hoeffding's Inequality:
\begin{theorem}

(Hoeffding's Inequality for general bounded random variables).

 Let $X_1,\dots,X_k$ be independent random variables. Assume that $X_t \in [m_t,M_t]$ almost surely for every $i$. Then, for any $\epsilon > 0$, we have \begin{align*}
 \p\left(\displaystyle\sum_{t=1}^{k} (X_t - \E[X_t]) \geq \epsilon\right) &\leq e^{{-\frac{2\epsilon^2}{\sum_{t=1}^{k}(M_t - m_t)^2}}}.
 \end{align*}
\end{theorem}

\begin{lemma}\label{lem:MoreCures}
The probability that the random walk reaches the stopping time with $I_{t_0} - I_t < 0$ tends to 0 as $\tau \to 0$.
\proof Using Hoeffding's Inequality: 

\begin{align*}  
\p \left(\sum_{t=1}^{k} A_t - B_t \geq 0 \right) &= \p \left.(\sum_{t=1}^{k} A_t - B_t - \E[A_t - B_t] \right. \geq  \left.- k\E[A_t - B_t] \right)  \\
&\leq \exp  \left(\frac{2\cdot ( k\E[A_t - B_t])^2}{\sum_{t=1}^{k} (r - (-\frac{r^3}{3}))^2} \right) \\
&\leq e^{-k  \frac{2(r\delta - \frac{r^3}{3}\mu)^2}{(r + \frac{r^3}{3})^2} }.
\end{align*}
Let $MoreCuring$ be the event that the increase in susceptible nodes at time $T_{\rm SmallCut}$ ($I_{t_0} - I_{T_{\rm SmallCut}}$)  is non-negative. We use Hoeffding's Inequality to bound $\p \left(\sum_{t=1}^{k} A_t - B_t \geq 0 \right)$. Then, b Proposition \ref{cl:GmanyInfection}, we know that at least $I = \frac{r^4}{7}$ infections must have taken place. To simplify the notations for this proof, we introduce two new stopping times, $T_{\rm Many Infections}$ and $T_{\rm Neg Binomial RW}$. $T_{\rm Small Cut}$ stochastically dominates $T_{\rm Many Infections}$, the number of time steps it takes for the random walk to infect $I$ new nodes. Since the infection rate is at least $\frac{r^3}{3}$,  $T_{\rm Many Infections}$ in turn stochastically dominates $T_{\rm Neg Binomial RW}$, a negative binomial distribution of parameter $\frac{3I}{r^3}$ and probability of failure $\mu$. We can therefore replace $T_{\rm Small Cut}$ by the simpler quantity $T_{\rm Neg Binomial RW}$ in the following calculations: 
\begin{align*}
\p(MoreCuring) &= \p \left(\sum_{t=1}^{T_{\rm Small Cut}} A_t - B_t \geq 0 \right) \\
&= \sum_{k=0}^{\infty} \p \left(\sum_{t=1}^{k} A_t - B_t \geq 0 \right) \cdot \p \left(T_{\rm Small Cut} = k \right) \\
&\leq \sum_{k=0}^{\infty} e^{-k  \frac{2(r^3\mu - r\delta)^2}{(r^3 + r)^2} } \cdot \p \left(T_{\rm Neg Binomial RW} = k \right) \\
&\leq e^{-\frac{I}{r^3}  \frac{6(r^3\mu - r\delta)^2}{(r^3 + r)^2} } \sum_{k=0}^{\infty} e^{-k  \frac{2(r^3\mu - r\delta)^2}{(r^3 + r)^2} } \cdot \p \left(T_{\rm Neg Binomial RW} = \frac{3I}{r^3} + k \right) \\
&\leq e^{-\frac{I}{r^3}  \frac{6(r^3\mu - r\delta)^2}{(r^3 + r)^2} } \left( \frac{\mu}{1-\mu e^{- \frac{2(r^3\mu - r\delta)^2}{(r^3 + r)^2} }} \right)^\frac{3I}{r^3} \\
&\to_{\tau \to 0} 0,
\end{align*}
where we have used that the MGF of a negative binomial of parameter M, probability of success $p$, evaluated at $u$, is $\left( \frac{1-p}{1-e^up} \right)^M $.
\qed
\end{lemma}

\subsection{There exists an infected node close to the root}  \label{sec:step2}

From the moment we start curing the last $\frac{N}{r^4}$ nodes, to the moment we have cured half of them and only $\frac{N}{2r^4}$ of these nodes remain infected, we show in this section that there exists an infected node at distance $\OO(\log\log(N))$ from the root (Proposition \ref{cl:rootclose}). This node stays infected for a high number of steps (Proposition \ref{cl:timeEndgame}). 

\begin{proposition} \label{cl:rootclose} 
If we select a set of $\frac{N}{2r^4}$ nodes in a tree such that the cut of this set is lower than $r^4$, then there is at least one node from this set at distance  $9\log(r) = 9\alpha\log\log(N) = \OO(\log\log(N))$ from the root.
\proof  We prove the contrapositive: if all the nodes of this set are at distance greater than $9\alpha\log\log(N)$ from the root, then the cut is higher than $r^4$.

Any subtree rooted at distance $9\alpha\log\log(N)$ from the root contains $\frac{N}{r^9}$ nodes, and has a cut of at least 1. Suppose all the $\frac{N}{2r^4}$ nodes of the selected set are at distance $9\alpha\log\log(N)$ or more from the root. We therefore need at least $\frac{N}{2r^4}/\frac{N}{r^9} = \frac{r^5}{2} $ such subtrees, for a total cut of at least $\frac{r^5}{2} > r^4 $.
Hence, the closest node is at distance at most $9\alpha\log\log(N) = \OO(\log\log(N))$ from the root. \qed 
\end{proposition}

We now show it takes many time steps to cure $\frac{N}{2r^4}$ nodes, regardless of the policy.

\begin{proposition} \label{cl:timeEndgame}
Curing half of the last $\frac{N}{r^4}$ nodes requires more than $ \frac{N}{2r^4} \cdot \frac{1}{\delta}$ time steps in expectation.
\proof If we ignore any potential infections, the time needed to cure $\frac{N}{2r^4}$ nodes is at least the sum of $\frac{N}{2r^4}$ geometric random variables of parameter $\delta$. The result follows by linearity of expectation. \qed 
\end{proposition}

\begin{proposition} 
Let $T_{\frac{N}{2r^4}}$ be the random variable representing the time to cure half of the $\frac{N}{r^4}$ last nodes. Then: 
$$\p\left(T_{\frac{N}{2r^4}} \leq \frac{N}{4r^5\delta}\right) \leq e^{-\frac{N}{8r^5}}.$$
\proof The proof can be found in the Appendix, Proposition \ref{cl:chernoffEndgame}.
\end{proposition}

Therefore, there exists an infected node close to exponentially many uninfected nodes, during at least $\frac{N}{4r^5 \delta}$ time steps. We now establish a lower bound on the probability of reinfecting $\frac{N}{r^4}$ new nodes in some other part of the graph, starting from this node.

\subsection{Low-cut case} \label{sec:BlindCuring}
We prove in this section that the probability of infecting $\frac{N}{r^4}$ new nodes in some other part of the graph, by the time it takes to cure half of the $\frac{N}{r^4}$ last infected nodes, is superpolynomially close to 1 for every strategy (Lemma \ref{lem:Pescape}). The graph can only be cured if this does not  happen.

The following Lemma is key to understanding why no strategy can prevent the reinfection. In the Blind Curing setting, we do not know which nodes are infected. Since there are exponentially many infection routes from the root of the tree, spreading the budget means there will always be a subtree on which very small budget is allocated. If the infection reaches this tree, reinfecting a lot of nodes becomes very likely.
\begin{lemma} \label{lem:minimalTree} 
For every time $t_0$, there exist $r$ subtrees containing $\frac{N}{r^3}$ nodes for which less than $\frac{t_3}{r}$ budget is used in the interval $[t_0, t_0 + t_3]$. We call any of these trees a \textbf{minimal tree} for $[t_0, t_0 + t_3]$.
\proof By the pigeonhole principle, since the total budget during this interval is $t_3 \cdot r$, and there are at least $r^3$ disjoint subtrees containing $\frac{N}{r^3}$ nodes (Proposition \ref{cl:Nsubtrees} in the appendix), there are at least $r$ subtrees that contain less than $ r \cdot \frac{t_3 \cdot r}{r^3} = \frac{t_3}{r}$ budget on this interval. \qed 
\end{lemma}

\begin{figure}[H]
\centering\includegraphics[width=8cm]{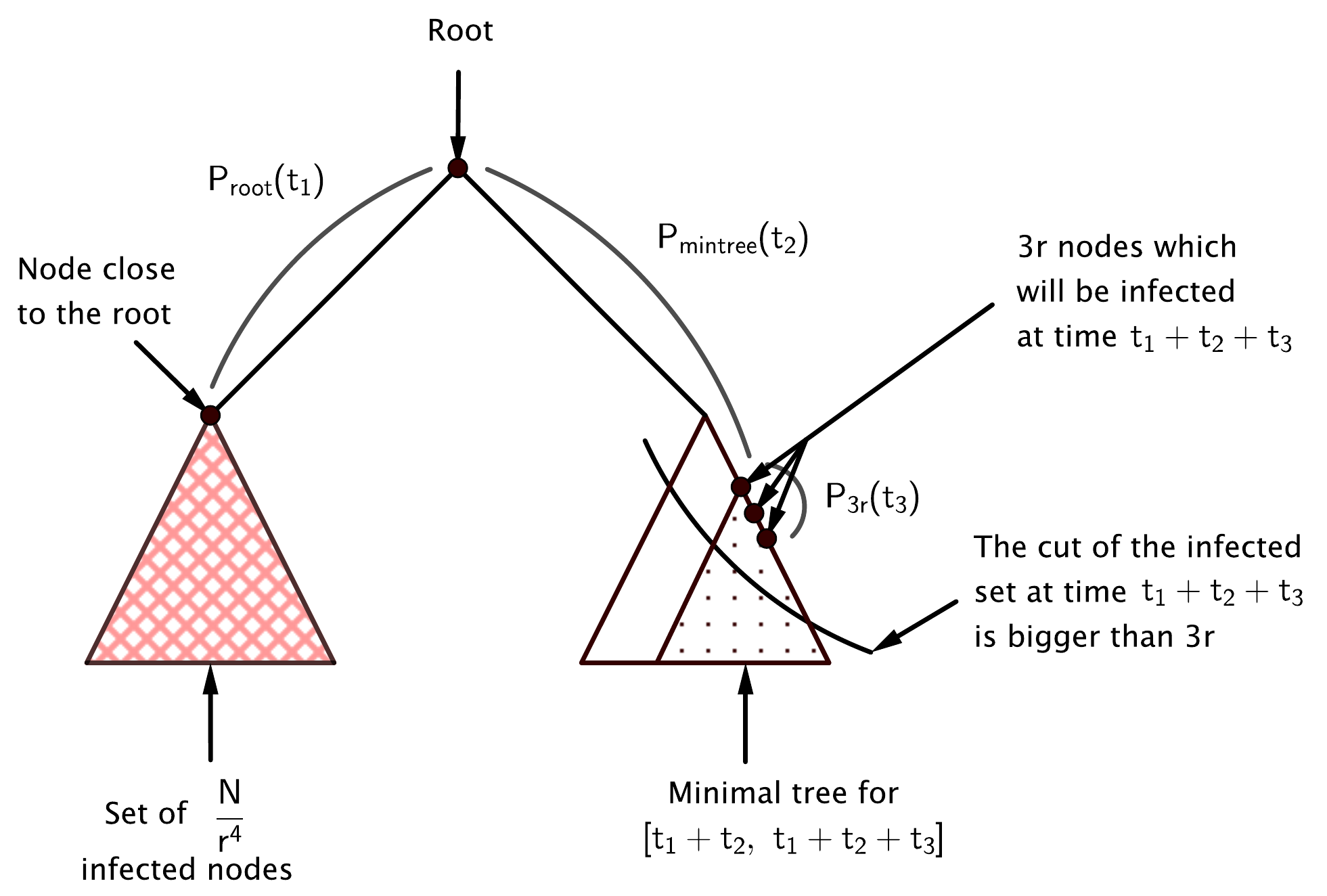}
\caption{Visual representation of $P_{\rm root}(t_1)$, $P_{\rm mintree}(t_2)$, and $P_{\rm 3r}(t_3)$.}
\end{figure}

\begin{definition}
From Proposition \ref{cl:rootclose}, we know there exists an infected node close to the root. We call an \textbf{Escape$(t_0,t_1, t_2, t_3)$} the conjunction of the following events:
\begin{enumerate}
	\item At time $t_0$, this node infects its parent.
	\item The infection propagates from the parent node to the root in time $t_1$, without any node being cured.
	\item The infection propagates from the root of the tree to the root of a minimal tree for $[t_0+t_1+t_2, t_0+t_1+t_2+t_3]$ in time $t_2$, without any node being cured.
	\item $3r$ new nodes are infected in a minimal tree for $[t_0+t_1+t_2, t_0+t_1+t_2+t_3]$ in time $t_3$, without any node in a minimal tree being cured.
	\item The number of newly infected nodes reaches $\frac{N}{r^4}$ before it reaches $r$.
\end{enumerate}
\end{definition}

We notice that if an $Escape$ happens, then $\frac{N}{r^4}$ new nodes in some other part of the graph were reinfected. However, it is possible to reinfect $\frac{N}{r^4}$ new nodes without any $Escape$ happening. 

If $\{t_0, t_1, t_2, t_3\} \neq \{t_0', t_1', t_2', t_3'\}$, then $Escape(t_0, t_1, t_2, t_3)$ and
\\ $Escape(t_0', t_1', t_2', t_3')$  are disjoint events.

We notice that the probability of all the events defined above is independent of $t_0$. To simplify notations, we set $t_0=0$ for the following definitions. 
\begin{itemize}
\item $P_{\rm root}(t_1)$, the probability that the infection reaches the root in time exactly $t_1$.
\item $P_{\rm mintree}(t_2)$, the probability that the infection reaches a minimal tree for $[t_1 + t_2, \, t_1 + t_2 + t_3]$ in time $t_2$, conditioned on the fact that the root of the tree is infected. Interestingly, by symmetry of the binary tree (all potential minimal trees are at the same distance from the root of the tree), this quantity does not depend on $t_1$ or $t_3$.
\item $P_{\rm 3r}(t_3)$, the probability that $3\cdot r$ nodes are reinfected in a minimal tree in time $t_3$, conditioned on the fact that the root of a minimal tree is infected. 
\item $P_{\rm spread}$, the probability that the increase in susceptible nodes since time $t_1 + t_2 + t_3$ reaches $-\frac{N}{r^4} + 3\cdot r$ before it reaches $2r$, conditioned on the fact that $3r$ new nodes are infected at time $t_1 + t_2 + t_3$.
\end{itemize}

The proof relies on the fact that \textit{no strategy can adapt to the infection moving towards a minimal tree}. In the Blind Curing setting, most of the budget is wasted covering nodes which are not infected or about to be infected, while most of the graph is left unprotected.

We now bound the probabilities defined above.

\begin{proposition}
\begin{align*} 
  \bullet \,\,P_{\rm root}(t_1) \,\quad&\geq   {t_1 \choose 9\alpha\log\log(N)}  \cdot \mu^{9\alpha\log\log(N)+1} (1-\mu)^{t_1 - 9\alpha\log\log(N)} (1-\delta)^{t_1}, \\
  \bullet \,\,P_{\rm mintree}(t_2) &\geq    {t_2 \choose 3\alpha\log\log(N)}    \cdot  \mu^{3\alpha\log\log(N)+1} (1-\mu)^{t_2 - 3\alpha\log\log(N)} (1-\delta)^{t_2}, \\
    \bullet \,\, P_{\rm 3r}(t_3) \,\,\,\,\,\quad&\geq  {t_3 \choose 3 r}  e^{-\frac{t_3}{r}\cdot \tau}  \cdot  \mu^{3 r+1} (1-\mu)^{t_3 - 3 r} e^{-\frac{t_3}{r}\cdot \tau}.
    \end{align*}
\proof 
\begin{itemize}
\item Straightforward combinatorics result.
\item Same, noticing $\log(r^3) = 3\alpha\log\log(N)$.
\item Since we are in a minimal tree for $[t_1 + t_2, \, t_1 + t_2 + t_3]$, the total budget that can be spread among all nodes during this time is $\frac{t_3}{r}$ (Lemma \ref{lem:minimalTree}). Let $r_i^{t}$ be the budget spent on node $i$ at time $t$, and let $\delta_i^{t}$ be the probability that node $i$ is cured at time $t$.
\begin{align*}
P_{\rm 3r}(t_3) &\geq   {t_3 \choose 3 r}  \mu^{3 r+1} (1-\mu)^{t_3 - 3 r} \prod_{\text{time} \, t=1}^{t_3} \prod_{\substack{\text{node } i \text{ in} \\ \text{minimal tree}}} (1 - \delta_i^{t} ) \\ 
&= {t_3 \choose 3 r}  \mu^{3 r+1} (1-\mu)^{t_3 - 3 r} \prod_{\text{time} \, t=1}^{t_3} \prod_{\substack{\text{node } i \text{ in} \\ \text{minimal tree}}} e^{-r_i^{t} \tau} \\
&= {t_3 \choose 3 r}  \mu^{3 r+1} (1-\mu)^{t_3 - 3 r}  e^{-\frac{t_3}{r}\cdot \tau}.
\end{align*}
\end{itemize}
\qed 
\end{proposition}

\begin{proposition} \label{cl:Pspread} 
Conditioned on the cut of the infected set being at least $3r$, the probability that the increase in susceptible nodes since time $t_1 + t_2 + t_3$ reaches $-\frac{N}{r^4} + 3\cdot r$ before it reaches $2r$, is at least $\frac{1 - \frac{1}{2}^{3\cdot r}}{1 - \frac{1}{2}^{\frac{N}{r^4}}} \geq \frac{1}{2}$.
\proof This is a classic Gambler's Ruin problem, with low boundary $2r$ and high boundary $\frac{N}{r^4}$. During the infection process, which starts with $3r$ infected nodes, the cut is always higher than $2r$, so the infection rate is always higher than $2r$, while the curing rate is $r$. The proof can be found in ~\cite{GrimmettGeoffreyStirzaker2001}.
\qed 
\end{proposition}

We now combine the previous results to bound the probability of escaping in one time step.

\begin{lemma} \label{lem:Pescape}
Let $P_{\rm escapeOneStep}$ be the probability that an $Escape$ starts at a given time step. Then:
\begin{align*}
P_{\rm escapeOneStep} &\geq \left(\mu \left(\frac{\mu (1-\delta)}{\delta + \mu (1-\delta)} \right)^{9\alpha\log\log(N)} \right) \cdot \left(\left(\frac{\mu e^{-\frac{1}{r}}}{(1-e^{-\frac{1}{r}}) + \mu (e^{-\frac{1}{r}})} \right)^{3r}\right)\cdot \left(\frac{1}{2} \right).
\end{align*}
Therefore, for $\tau$ sufficiently small (and in particular, as $\tau \to 0$),
$$P_{\rm escapeOneStep} \geq  \frac{\tau}{2e^3e^{12\alpha^2\log^2\log(N)}} +o(\tau).  $$
\proof We notice $P_{\rm root}(t_1)$ only depends on $t_1$, $P_{\rm mintree}(t_2)$ only depends on $t_2$, $P_{\rm 3r}(t_3)$ only depends on $t_3$, while $P_{\rm spread}$ is independent of $t_1$, $t_2$, and $t_3$. Therefore, using results from Lemma \ref{lem:Ppath} and Corollary \ref{cor:Ppath} in the Appendix which give the values of $P^{\rm startPath}$ and $P^{\rm path}$, we have:

\begin{align*}
P_{\rm escapeOneStep} &= \displaystyle\sum_{t_1, t_2, t_3 = 0}^{\infty} P_{\rm root}(t_1) \cdot  P_{\rm mintree}(t_2)  \cdot  P_{\rm 3r}(t_3) \cdot  P_{\rm spread}  \\
&= \left(\sum_{t_1= 0}^{\infty} P_{\rm root}(t_1) \right)\cdot \left(\sum_{t_2= 0}^{\infty} P_{\rm mintree}(t_2) \right) \cdot \left(\sum_{t_3 = 0}^{\infty}  P_{\rm 3r}(t_3)\right)\cdot \left(P_{\rm spread} \right) \\
&\geq \left(P^{\rm startPath}_{9\alpha\log\log(N)} \right)\cdot \left(P^{\rm path}_{3\alpha\log\log(N)} \right)\cdot \left(\left(\frac{\mu e^{-\frac{\tau}{r}}}{(1-e^{-\frac{\tau}{r}}) + \mu (e^{-\frac{\tau}{r}})} \right)^{3r}\right) \cdot \left(P_{\rm spread} \right) \\
&\geq \left(\mu\left(\frac{\mu (1-\delta)}{\delta + \mu (1-\delta)} \right)^{12\alpha\log\log(N)} \right)\cdot \left(\left(\frac{\mu e^{-\frac{\tau}{r}}}{(1-e^{-\frac{\tau}{r}}) + \mu (e^{-\frac{\tau}{r}})} \right)^{3r}\right)\cdot \left(\frac{1}{2} \right).
\end{align*}

 As $\tau \to 0$:

\begin{align*}
P_{\rm escapeOneStep} &\sim_{\tau \to 0}  \left(\tau \left(\frac{\tau }{(r+1)\tau} \right)^{12\alpha\log\log(N)} \right)\cdot \left(\frac{\tau}{(\frac{1}{r} + 1) \tau} \right)^{3r}\cdot \frac{1}{2}  \\
&\geq_{\tau \to 0} \tau  \left(e^{-12\alpha^2\log^2\log(N)} \right)\cdot \frac{e^{-\log(1+ \frac{1}{r})\cdot 3r}}{2} +o(\tau) \\
&\geq_{\tau \to 0} \tau  \left(e^{-12\alpha^2\log^2\log(N)} \right) \cdot \frac{e^{-3}}{2} +o(\tau) \\
&\geq_{\tau \to 0} \frac{\tau}{2e^3e^{12\alpha^2\log^2\log(N)}} +o(\tau).
\end{align*} 
\qed 
\end{lemma}

We therefore deduce the probability that no $Escape$ happens by the time we cure half of the $\frac{N}{r^4}$ infected nodes.

\begin{lemma} \label{lem:pNoescape}
Let $NoEscape$ be the event that no $Escape$ happens by the time we cure half of the $\frac{N}{r^4}$ infected nodes. Then:

\[\p \left(NoEscape \right) \leq_{N \gg 1} e^{-\frac{N}{e^{24\alpha^2\log^2\log(N)}}}. \]
\proof Since there are always more than $\frac{N}{2r^4}$ nodes infected, there is at least one infected node at distance $9\alpha\log\log(N)$ from the root (Lemma \ref{cl:rootclose}), which means the bound for $P_{\rm escapeOneStep}$ established in Lemma \ref{lem:Pescape} holds. Using Proposition \ref{cl:chernoffEndgame}, we split the analysis into two cases: whether we can cure  $\frac{N}{2r^4}$ in less than $\frac{N}{4r^5\delta}$ time steps or not. The probability of one $Escape$ starting at time $t$ being independent from the probability of an $Escape$ starting at any other time step $t'$:
\begin{align*}
\p \left(NoEscape \right) &\leq  \left(1- P_{\rm escapeOneStep} \right)^{T_{\frac{N}{2r^4}}} \\
&\leq  \p(T_{\frac{N}{2r^4}} \leq \frac{N}{4r^5\delta})\cdot  1 + \p(T_{\frac{N}{2r^4}} \geq \frac{N}{4r^5\delta})\cdot \left(1- P_{\rm escapeOneStep} \right)^{\frac{N}{4r^5\delta}} \\
&\leq  e^{-\frac{N}{8r^5}} + \left(1- P_{\rm escapeOneStep} \right)^{\frac{N}{4r^5\delta}}.
\end{align*}

Using Lemma \ref{lem:Pescape} to get an equivalent when $\tau \to 0$:

\begin{align*}
\p \left(NoEscape \right) &\leq e^{-\frac{N}{8r^5}} +  \left(1- \left(\mu\frac{\mu (1-\delta)}{\delta + \mu (1-\delta)} \right)^{12\alpha\log\log(N)} \right)^{\frac{N}{4 r^5\delta}} \\
&\leq_{\tau \to 0} e^{-\frac{N}{8r^5}} + \left(1 - \frac{\tau}{2e^3e^{12\alpha^2\log^2\log(N)}} + o(\tau) \right)^{\frac{N}{2r^5r\tau}}  \\ 
&\leq_{N \gg 1} e^{-\frac{N\tau}{4e^3r^6e^{12\log^2(r)}\tau}} \\
&\leq_{N \gg 1} e^{-\frac{N}{e^{24\alpha^2\log^2\log(N)}}}.
\end{align*}
\qed
\end{lemma}

\subsection{A Blind Curing result}
From Sections \ref{sec:step1} and \ref{sec:BlindCuring}, we know the graph can only be cured if we are in one of these two cases:
\begin{enumerate}
	\item The cut was above $r^4$, but we cured the whole graph anyway, which happens with probability less than $e^{-\log^2(N)}$ (Proposition \ref{cor:polylog})
	\item The cut was below $r^4$, but no $Escape$ happens by the time it takes to cure half of $\frac{N}{r^4}$ infected nodes, which happens with probability less than $e^{-\frac{N}{e^{24\alpha^2\log^2\log(N)}}}$ (Lemma \ref{lem:pNoescape}). 
\end{enumerate}

We can therefore obtain a bound on the expected time it takes to cure the whole graph.

\begin{theorem} \label{th:BlindCuring} 
In the Blind Curing setting, curing a complete binary tree takes $\Omega \left(   e^{\log^2(N)}  \right)  $ time in expectation with any budget polynomial in the {\sc CutWidth}. Therefore, no polynomial expected time curing strategy exists for budget $r=\OO(W^\alpha) = \OO(\log^\alpha(N))$, for all $\alpha$ constant.
\proof Let $CureLastNodes$ be the event that we are in case (1) or (2) described above. By union bound:
\begin{align*}
\p \left(CureLastNodes\right) &\leq e^{-\log^2(N)} + e^{-\frac{N}{e^{24\alpha^2\log^2\log(N)}}} \\
	&\leq 2e^{-\log^2(N)}.
\end{align*}
Using Proposition \ref{cl:taudelta} of the appendix, we have $\frac{1}{r} \leq \frac{\tau}{\delta}$. The number of times we try to cure the last $\frac{N}{r^4}$ is stochastically bounded below by a geometric variable of parameter $\p \left(CureLastNodes\right)$. Following Proposition \ref{cl:timeEndgame}, curing $\frac{N}{2r^4}$ lasts at least $\frac{N}{2r^4} \cdot \frac{1}{\delta}$ time steps, so $\frac{N}{2r^4}\cdot \frac{\tau}{\delta}$ time. Therefore, the expectation of the length of the curing process is the number of times we try to cure the last infected nodes, multiplied by the time it takes to cure them.

\begin{align*}
\mathbb{E}(\text{Length}) 
&\geq \underbrace{\frac{1}{\p \left(CureLastNodes\right) }}_{\substack{\text{expected number} \\ \text{of times we try}  \\ \text{to cure $\frac{N}{r^4}$ nodes}}} \cdot  \underbrace{\frac{N}{2r^4\delta}}_{\substack{\text{minimal number of } \\ \text{time steps to}  \\ \text{cure $\frac{N}{r^4}$ nodes}}} \cdot \quad \underbrace{\tau}_{\substack{\text{size of a} \\ \text{time step}}} \\
&\geq \frac{e^{\log^2(N)}}{2}  \cdot \frac{N}{2r^4} \cdot \frac{1}{r} \\
&= \Omega \left( e^{\log^2(N)} \right).
\end{align*} 
Hence, it is not curable in polynomial time for budget $r=\OO(W^\alpha) = \OO(\log^\alpha(N))$, for all $\alpha$ constant. \qed 
\end{theorem}

	\section{Proof for Partial Information} \label{sec:partial}
\begin{definition}
We call \textbf{sample} the information given by one node at a given time step (i.e., whether a flag was raised or not). We call an \textbf{infected-sample} a sample from an infected node.
\end{definition}

When an $Escape$ happens, we show that with constant probability, not too many infected-samples are produced. In particular, by the time reinfecting $\frac{N}{r^4}$ new nodes becomes inevitable with constant probability, not enough infected-samples are produced to determine if one of the minimal trees' nodes is infected with better than constant error probability. In other words, no strategy can utilize the information available without making mistakes a constant fraction $P_{confuse}$ of the time.

If we cannot recognize that an infection has happened before it is too late to prevent it, everything is as if we were in the Blind Curing model. We can therefore extend the results from the previous section.

\subsection{Quantity of information available before the cut reaches $3r$}

If we reuse the terms introduced in Section \ref{sec:BlindCuring}, the $Escapes$ we consider are composed of four phases: 
\begin{enumerate}
\item reaching the root
\item reaching the root of a minimal tree. There are $r$ possible such minimal trees.
\item infecting $3r$ nodes in this minimal tree
\item spreading the infection from $3r$ to $\frac{N}{r^4}$ nodes
\end{enumerate}
Since the spreading phase (4) happens with constant probability even in the Complete Information setting, we focus on the number of samples created by the first three phases. We focus in particular on the number of samples created by phase (3) in Lemma \ref{lem:nSamples}. We show in the proof of Lemma \ref{lem:escapeTime} that the number of samples produced by phases (1) and (2) is negligible compared to the number of samples produced by phase (3). To make sure that no other infected-samples can be gathered, we forbid infections from happening outside of an $Escape$ (Lemma \ref{cl:noOtherInfection}).

Let us notice that every time a new infection takes place, the cut increases by 1 (every new node infected gives access to 2 new nodes, but the edge leading to it is not part of the cut any longer).

The next lemma says that in the event of an infection in a minimal tree, it is likely that the number of infected-samples we obtain is small.

\begin{lemma}\label{lem:nSamples} 
In the event that the root of the minimal tree becomes infected, then conditioned on the event that $3r$ new nodes of this minimal tree become infected, we gather at most $\frac{6r}{\tau}$ samples from the newly infected nodes, with probability at least $\frac{1}{2}$.
\proof Let $N_{\rm samples}$ be the number of samples produced by infected nodes from the time one node was infected to the moment the $3r^{th}$ node was infected. The time to infect one more node when $i$ nodes are infected is given by a geometric variable $Geo(i,\mu)$ of parameter $1-(1-\mu)^i$ (Proposition \ref{cl:minGeo} of the appendix). Therefore, the $j$th node to be infected produces $\sum_{i=j}^{3r-1} Geo(i,\mu)$ samples. Thus, conditioned on the event that $3r$ nodes become infected:
\begin{align*}
N_{\rm samples} &= \sum_{j=1}^{3r-1} \sum_{i=j}^{3r-1} Geo(i,\mu) = \sum_{j=1}^{3r-1} j\cdot Geo(j,\mu) .
\end{align*}
Therefore, again conditioned on $3r$ nodes becoming infected, the expected number of samples is:
\begin{align*}
\E(N_{\rm samples} \,| \, \mbox{$3r$ infected}) &= \sum_{j=1}^{3r-1} j\cdot \E(Geo(j,\mu)) = \sum_{j=1}^{3r-1} j\cdot \frac{1}{1-(1-\mu)^j} \\
&=_{\tau \to 0} \sum_{j=1}^{3r-1} j\cdot  \frac{1}{j\tau} \qquad \leq_{\tau \to 0} \frac{3r}{\tau}.
\end{align*}
We conclude by using Markov's Inequality.
\qed
\end{lemma}

We now count the number of samples available from phases (1), (2) and (3) of an $Escape$.

\begin{lemma} \label{lem:escapeTime} 
Conditioned on the event that an $Escape$ happened, we gather at most  $\frac{6r}{\tau}$ infected-samples from phase (3), and $\frac{360\log^2(r)}{\tau}$ infected-samples from phases (1) and (2), with probability at least $\frac{1}{4}$.
\proof
Using Proposition \ref{cl:rootclose}, there are at most $9\log(r)$ nodes from one infected node to the root. Using Proposition \ref{lem:minimalTree}, the minimal tree is at distance $3\log(r)$ from the root. We then need $3r$ additional infections to get to a point where the infection is unstoppable (Proposition \ref{cl:Pspread}).  Using Markov's Inequality, we can infect these $9\log(r) + 3\log(r) = 12 \log(r)$ nodes in $\frac{24\log(r)}{\tau}$ time steps with probability $\frac{1}{2}$. Using Lemma \ref{lem:infectionTime} of the appendix, we can infect the $3r$ nodes in $\frac{2\log(3r)}{\tau} \leq \frac{6\log(r)}{\tau}$ time steps with probability $\frac{1}{2}$. Therefore, we can infect all these nodes in $\frac{30\log(r)}{\tau}$ time steps with probability $\frac{1}{4}$, which gives at most $\frac{30\log(r)}{\tau} \cdot 12 \log(r)$ samples for the first $12 \log(r)$ nodes, and $2 \cdot \frac{3r}{\tau} = \frac{6r}{\tau}$ samples for the last $3r$ nodes, which concludes the proof.
\qed
\end{lemma}

Conditioned on reaching a cut of $3r$ in a minimal tree in less than $\frac{30\log(r)}{\tau}$ time steps, we now bound the probability of not infecting any nodes which are not part of the $Escape$. This ensures that the only infected samples we could get come from the nodes in the $Escape$.

\begin{proposition}
	Conditioned on reaching a cut of $3r$ in a minimal tree in less than $\frac{30\log(r)}{\tau}$ time steps, the probability $P_{\rm NoOtherInfections}$ of not infecting any nodes outside of the $Escape$  is bounded by:
		\begin{align*}
		 P_{\rm NoOtherInfections} &\leq e^{-\frac{360\log^2(r)\mu}{\tau}} \\
		 &\leq_{\tau \to 0} e^{-360\log^2(r)}.
		 \end{align*}
	\proof The proof can be found in Proposition  \ref{cl:noOtherInfection} in the Appendix.
\end{proposition}

\subsection{Distinguishing between infected and not infected}
The proof relies on this idea: any strategy attempting to prevent an $Escape$ needs to shift its budget towards the minimal tree in which the infection is progressing. For this to happen, it is necessary to realize that one of the minimal trees is threatened. Determining which one amounts to distinguishing between the following hypotheses: 
\begin{itemize}
\item $H_0$: In the null hypothesis, none of the $r$ minimal trees have any infected nodes.
\item $H_1$: One of the $r$ minimal trees has at least one infected node, while the others do not.
\end{itemize}

We use the results in Lemma \ref{cl:samples} to show that there are not enough infected-samples created by nodes on the path from the node close to the root to the root of a minimal tree to realize that even one node is infected. This implies that the nodes from phase (1) and (2) cannot help detect a threat to a minimal tree. Lemma  \ref{cl:samples} is also used to show that we do not gather enough infected-samples from phase (3) to distinguish between $H_0$ and $H_1$ defined above, which means we cannot know if there is at least one infected node in one minimal tree.

Thus, we combine all these results to calculate $\p(NoEscapePI)$, the probability that an $Escape$ happens, that not too many samples are produced during this $Escape$, that no other nodes are infected outside of the $Escape$, that the samples from phase (1) and (2) do not allow the identification of the infected minimal tree, and that the samples from phase (3) are not enough to reveal whether or not one minimal tree is indeed infected. 

We finally use these results to extend the Blind Curing theorem to the Partial Information setting (Theorem \ref{th:partialInfo}).

\begin{lemma} \label{cl:samples} 
\begin{enumerate}
\item We need $\Omega \left(  \frac{\log\left(\frac{1}{\epsilon}\right)}{\mathcal{D}(p||q)} \right)$ samples to decide if a node is infected or not with probability at least $\Omega \left( 1 - \epsilon \right)$.
\item We need $\Omega \left(  \frac{\log\left(\frac{1}{\epsilon}\right) \sqrt{\log(r)}}{\mathcal{D}(p||q)} \right)$ samples to distinguish between hypothesis $H_0$ and $H_1$, and detect if one minimal tree has at least one node infected among $r$ minimal trees.
\end{enumerate}

\proof \begin{enumerate}
\item Using Sanov's Theorem ~\cite{Sanov1961}, following the proof in Proposition 5.6 of ~\cite{Mansour2011}, we know that we need $\Omega\left(\frac{1}{\mathcal{D}(p||q)}\right)$ samples to distinguish between two coins of parameters $p$ and $q$ with probability $\Omega(1)$. We can boost this probability to show that we need  $\Omega \left(  \frac{\log\left(\frac{1}{\epsilon}\right)}{\mathcal{D}(p||q)} \right)$ to distinguish between $p$ and $q$ with probability $\Omega \left( 1 - \epsilon \right)$.
\item This follows by considering the maximum of $r$ $\mathcal{B}(n,q)$ binomial random variables. Using a Gaussian approximation to the binomial, and the fact that the maximum of $r$ standard Gaussian random variables is $\sqrt{2 \log(r)}$, we obtain the desired result. 
\end{enumerate}
\qed
\end{lemma}

\begin{corollary} \label{cor:Pconfuse} 
The probability $P_{\rm confuse}$ of not being able to detect which minimal tree is infected during phase (3) is bounded away from 0. 

In particular, we have:
\begin{align*}
P_{\rm confuse} &=_{N \gg 1} \Omega\left(e^{-\frac{\mathcal{D}(p||q)}{\tau} \cdot \frac{r}{\sqrt{\log(r)}} } \right).
\end{align*}
\proof We separate the samples in two groups:
\begin{itemize}
	\item The sample from phase (1) and (2) are used to detect if one node was infected on the path from the node close to the root to the root of a minimal tree. We get $\frac{360\log^2(r)}{\tau} = \frac{\log\left(e^{\frac{\mathcal{D}(p||q)}{\tau} \cdot 360\log^2(r) }\right) } {\mathcal{D}(p||q)}$ samples from phase (1) and (2) (Lemma \ref{lem:nSamples}), so the probability of confusing coins of parameters $p$ and $q$ is at least $ \Omega\left(e^{-\frac{\mathcal{D}(p||q)}{\tau} \cdot 360\log^2(r) } \right)$ according to Lemma \ref{cl:samples}.
	\item The samples from all phases are used to distinguish between $H_0$ and $H_1$. We get $\frac{6r}{\tau} = \frac{\log\left(e^{\frac{\mathcal{D}(p||q)}{\tau} \cdot \frac{6r}{\sqrt{\log(r)}} }\right) \sqrt{\log(r)} } {\mathcal{D}(p||q)}$ samples from phase (1) and (2) (Lemma \ref{lem:nSamples}), so the probability of confusing coins of parameters $p$ and $q$ is at least $ \Omega\left(e^{-\frac{\mathcal{D}(p||q)}{\tau} \cdot \frac{6r}{\sqrt{\log(r)} } } \right)$ according to Lemma \ref{cl:samples}.
\end{itemize}
Combining the two, the probability of not detecting the threat is at least:
\begin{align*}
	P_{\rm confuse} &= \Omega\left(e^{-\frac{\mathcal{D}(p||q)}{\tau} \cdot (\frac{6r}{\sqrt{\log(r)} } + 360\log^2(r))} \right) \\
	&=_{N \gg 1} \Omega\left(e^{-\frac{\mathcal{D}(p||q)}{\tau} \cdot \frac{r}{\sqrt{\log(r)} } } \right).
\end{align*}
\qed
\end{corollary}
We now consider the time needed to cure the graph for  $P_{\rm confuse} > 0$.

\begin{lemma} \label{lem:timePconfuse} 
Let $EscapePI$ be the event that by the time it takes to cure half of $\frac{N}{r^4}$ infected nodes, an $Escape$ happens but remains undetectable (\textit{i.e.,} the samples produced by the newly infected nodes during phases (1), (2), and (3) are not enough to deduce that there exists an infected minimal tree), and no node outside of the $Escape$ becomes infected. We provide a bound for $NoEscapePI$, the complementary of this event.

$$ \p(NoEscapePI) \leq e^{-\frac{P_{\rm NoOtherInfections} \cdot P_{\rm confuse} \cdot N}{e^{96\alpha^2\log^2\log(N)}}}.$$
\proof Let $EscapeOneStepPI$ be the conjunction of all the following events:
\begin{itemize}
	\item An $Escape$ happens at a given time step, which happens with probability at least $P_{\rm EscapeOneStep}$ (Lemma \ref{lem:Pescape}).
	\item Conditioned on an $Escape$ happening, less than $\frac{6r}{\tau}$ samples are produced by the newly infected nodes during phase (3), and less than $\frac{360\log^2(r)}{\tau}$ infected-samples are produced during phase (1)-(2), which happens with probability at least $\frac{1}{4}$ (Lemma \ref{lem:nSamples}).
	\item Conditioned on an $Escape$ happening in less than $\frac{30\log(r)}{\tau}$ time steps, no node outside of the $Escape$ becomes infected, which happens with probability $P_{\rm NoOtherInfections}$.
    \item Conditioned on all the above, the samples from phase (3) are not enough to reveal whether or not one minimal tree is indeed infected, which happens with probability $P_{\rm confuse}$.
\end{itemize} 
We notice that if we cannot tell whether or not a minimal tree is infected by the time it takes to reach phase (4) of an $Escape$, the situation is almost equivalent to the Blind Curing model. We can therefore apply exactly the same reasoning as in Theorem \ref{th:BlindCuring} if we replace $\p(EscapeOneStep)$ by $\p(EscapeOneStepPI)$. 
\begin{align*} 
\p(EscapeOneStepPI) \geq &P_{\rm EscapeOneStep} \cdot \frac{1}{4}  \\
&\, \cdot P_{\rm NoOtherInfections} \cdot P_{\rm confuse},
 \end{align*}
Following the exact same reasoning as in Lemma \ref{lem:Pescape}, we get:
$$ \p(NoEscapePI) \leq e^{-\frac{P_{\rm NoOtherInfections} \cdot P_{\rm confuse} \cdot N}{e^{96\alpha^2\log^2\log(N)}}}.$$
\qed
\end{lemma}

\begin{theorem}{A Partial Information impossibility result} \label{th:partialInfo} \\
Let $\frac{\mathcal{D}(p||q)}{\tau}$ be a measure of the amount of information we get by time step. If:
\begin{align*}
\frac{\mathcal{D}(p||q)}{\tau} &\leq \OO\left( \left(\log\left(\frac{N}{e^{456\alpha^2\log^2\log(N)}} \right)  - 2\log\log(N)\right)\frac{\sqrt{\log(r)} }{r} \right)  \\ 
&= \OO\left(\frac{\log(N)\sqrt{\log(r)} }{r}\right),
\end{align*}  as $\tau \to 0$, we cannot cure the complete binary tree in polynomial expected time with budget $r = W^\alpha$, for any $\alpha$ constant.
\proof 
From Lemma \ref{lem:timePconfuse}, we know: 
$$ \p(NoEscapePI) \leq e^{-\frac{P_{\rm confuse} \cdot P_{\rm NoOtherInfections} \cdot N}{e^{96\alpha^2\log^2\log(N)}}}.$$
From Corollary \ref{cor:Pconfuse}, we know 
$$P_{\rm confuse} \geq \Omega\left(e^{-\frac{\mathcal{D}(p||q)}{\tau} \cdot \frac{r}{\sqrt{\log(r)} } } \right).$$
\begin{align*}
 \p(NoEscapePI) &\leq e^{-\frac{ P_{\rm NoOtherInfections} \cdot P_{\rm confuse} \cdot N}{e^{96\alpha^2\log^2\log(N)}}} \\
 &\geq  e^{-\OO\left(\frac{ e^{- \cdot\left(\log\left(\frac{N}{e^{456\alpha^2\log^2\log(N)}} \right)  - 2\log\log(N)\right)  \cdot \frac{r\sqrt{\log(r)}}{\sqrt{\log(r)} r}} \cdot N}{e^{456\alpha^2\log^2\log(N)}}\right)} \\
 &\geq e^{-\OO\left(\log^2(N)\right)}.
\end{align*}
Following the same reasoning as in Theorem \ref{th:BlindCuring}, we conclude it takes at least $ \frac{e^{\Omega \left(  \log^2(N)\right)} \cdot N}{2\log^{4\alpha}(N)}  \geq e^{\Omega\left(  \log^2(N)\right)} $ time to cure the graph, so more than any polynomial expected time.
\qed 
\end{theorem}
In particular, this holds for $\alpha = 1$. If we remember that the {\sc CutWidth} of a tree is smaller than $\log(N)$ (Proposition \ref{cl:Wtree} of the appendix), we obtain:

\begin{corollary}
If the quantity of information by time step measured by $\frac{\mathcal{D}(p||q)}{\tau} $ is constant, no strategy can achieve polynomial time curing for the complete binary tree in the Partial Information setting, for budget $r= \OO(W) = \OO(\log(N))$.
\end{corollary}

	\section{Conclusion}
We have shown that unless we know the state of each node with perfect accuracy, and instantaneously, then the {\sc CutWidth} of the graph is no longer the sole quantity which determines the budget required to cure an infection in polynomial time. Practically, this means that quickly obtaining signature-based diagnostic tools, even if expensive, is critical. On the theoretical side, our work shows that the interplay between stochastic processes and combinatorial properties of graphs needs to be better understood. Indeed, resolving the gap between our upper and lower bounds as a function of general topological graph quantities remains an important question. Similarly, extending our understanding of upper and lower bounds to other infection models is important. This work demonstrates the important connection between budget for control, and budget for estimation, as for many interesting problems, these two are inextricably intertwined. 
	\newpage
	\newpage
	\balance
	\bibliographystyle{plain}
	\bibliography{Mendeley,const} 

\begin{thebibliography}{10}

\bibitem{Arias-castro2011}
Ery Arias-castro, Emmanuel~J Cand{\`{e}}s, and Arnaud Durand.
\newblock {Detection of an anomalous cluster in a network}.
\newblock 39(1):278--304, 2011.

\bibitem{Arias-Castro2008}
Ery Arias-Castro, Emmanuel~J. Cand{\`{e}}s, Hannes Helgason, and Ofer Zeitouni.
\newblock {Searching for a trail of evidence in a maze}.
\newblock {\em Annals of Statistics}, 36(4):1726--1757, 2008.

\bibitem{Arias-castro}
Ery Arias-castro and S~T Nov.
\newblock {Detecting a Path of Correlations in a Network}.
\newblock pages 1--12.

\bibitem{Bernoulli2004}
Daniel Bernoulli and Sally Blower.
\newblock {An attempt at a new analysis of the mortality caused by smallpox and
  of the advantages of inoculation to prevent it}.
\newblock {\em Reviews in medical virology}, 14:275--288, 2004.

\bibitem{bose2008}
Abhijit Bose, Xin Hu, Kang~G. Shin, and Taejoon Park.
\newblock Behavioral detection of malware on mobile handsets.
\newblock In {\em Proceedings of the 6th International Conference on Mobile
  Systems, Applications, and Services}, MobiSys '08, pages 225--238, New York,
  NY, USA, 2008. ACM.

\bibitem{Brown}
Daniel~G Brown.
\newblock How i wasted too long finding a concentration inequality for sums of
  geometric variables.
\newblock {\em Found at https://cs. uwaterloo. ca/\~{} browndg/negbin. pdf}, 6.

\bibitem{cover2012elements}
Thomas~M Cover and Joy~A Thomas.
\newblock {\em Elements of information theory}.
\newblock John Wiley \& Sons, 2012.

\bibitem{Drakopoulos2014}
Kimon Drakopoulos, Asuman Ozdaglar, and John~N. Tsitsiklis.
\newblock {An efficient curing policy for epidemics on graphs}.
\newblock {\em arXiv preprint arXiv:1407.2241}, (December):1--10, 2014.

\bibitem{Drakopoulos2015}
Kimon Drakopoulos, Asuman Ozdaglar, and John~N. Tsitsiklis.
\newblock {A lower bound on the performance of dynamic curing policies for
  epidemics on graphs}.
\newblock (978):3560--3567, 2015.

\bibitem{Drakopoulos2015a}
Kimon Drakopoulos, Asuman Ozdaglar, and John~N. Tsitsiklis.
\newblock {When is a network epidemic hard to eliminate?}
\newblock pages 1--17, 2015.

\bibitem{fanti2016rumor}
Giulia Fanti, Peter Kairouz, Sewoong Oh, Kannan Ramchandran, and Pramod
  Viswanath.
\newblock Rumor source obfuscation on irregular trees.
\newblock In {\em Proceedings of the 2016 ACM SIGMETRICS International
  Conference on Measurement and Modeling of Computer Science}, pages 153--164.
  ACM, 2016.

\bibitem{fanti2015spy}
Giulia Fanti, Peter Kairouz, Sewoong Oh, and Pramod Viswanath.
\newblock Spy vs. spy: Rumor source obfuscation.
\newblock In {\em ACM SIGMETRICS Performance Evaluation Review}, volume~43,
  pages 271--284. ACM, 2015.

\bibitem{Gallager2013}
Robert Gallager.
\newblock {Stochastic Processes: 9 - Random Walks, Large Deviations, and
  Martingales}.
\newblock {\em Stochastic Processes}, 2013.

\bibitem{ganesh2005effect}
Ayalvadi Ganesh, Laurent Massouli{\'e}, and Don Towsley.
\newblock The effect of network topology on the spread of epidemics.
\newblock In {\em INFOCOM 2005. 24th Annual Joint Conference of the IEEE
  Computer and Communications Societies. Proceedings IEEE}, volume~2, pages
  1455--1466. IEEE, 2005.

\bibitem{GrimmettGeoffreyStirzaker2001}
David {Grimmett, Geoffrey Stirzaker}.
\newblock {\em {Probability and random processes}}.
\newblock Oxford university press, 2001.

\bibitem{gregoire2008}
Grégoire Jacob, Hervé Debar, and Eric Filiol.
\newblock Behavioral detection of malware: from a survey towards an established
  taxonomy.
\newblock {\em Journal in Computer Virology}, 4(3):251--266, 2008.

\bibitem{Mansour2011}
Yishay Mansour.
\newblock {Lecture 5 : Lower Bounds using Information Theory Tools Distance
  between Distributions KL-Divergence}.
\newblock 2011.

\bibitem{meirom2015}
Eli~A Meirom, Chris Milling, Constantine Caramanis, Shie Mannor, Sanjay
  Shakkottai, and Ariel Orda.
\newblock Localized epidemic detection in networks with overwhelming noise.
\newblock In {\em ACM SIGMETRICS Performance Evaluation Review}, volume~43,
  pages 441--442. ACM, 2015.

\bibitem{Milling2015a}
Chris Milling, Constantine Caramanis, Shie Mannor, and Sanjay Shakkottai.
\newblock {Distinguishing Infections on Different Graph Topologies}.
\newblock {\em IEEE Transactions on Information Theory}, 61(6):3100--3120,
  2015.

\bibitem{Milling2015}
Chris Milling, Constantine Caramanis, Shie Mannor, and Sanjay Shakkottai.
\newblock {Local detection of infections in heterogeneous networks}.
\newblock {\em Proceedings - IEEE INFOCOM}, 26:1517--1525, 2015.

\bibitem{Newman2002}
Mark E.~J. Newman.
\newblock {Spread of epidemic disease on networks}.
\newblock {\em Physical Review E - Statistical, Nonlinear, and Soft Matter
  Physics}, 66(1), 2002.

\bibitem{Sanov1961}
Ivan~N. Sanov.
\newblock {On the Probability of Large Deviations of Random Variables}, 1961.

\bibitem{shah2010detecting}
Devavrat Shah and Tauhid Zaman.
\newblock Detecting sources of computer viruses in networks: theory and
  experiment.
\newblock In {\em ACM SIGMETRICS Performance Evaluation Review}, volume~38,
  pages 203--214. ACM, 2010.

\bibitem{shah2011rumors}
Devavrat Shah and Tauhid Zaman.
\newblock Rumors in a network: Who's the culprit?
\newblock {\em IEEE Transactions on information theory}, 57(8):5163--5181,
  2011.

\bibitem{shah2012rumor}
Devavrat Shah and Tauhid Zaman.
\newblock Rumor centrality: a universal source detector.
\newblock In {\em ACM SIGMETRICS Performance Evaluation Review}, volume~40,
  pages 199--210. ACM, 2012.

\bibitem{Sharpnack2012}
James Sharpnack, Alessandro Rinaldo, and Aarti Singh.
\newblock {Changepoint detection over graphs with the spectral scan statistic}.
\newblock {\em arXiv preprint}, 31:1--14, 2012.

\bibitem{spencer2015impossibility}
Sam Spencer and R~Srikant.
\newblock On the impossibility of localizing multiple rumor sources in a line
  graph.
\newblock {\em ACM SIGMETRICS Performance Evaluation Review}, 43(2):66--68,
  2015.

\bibitem{wang2014rumor}
Zhaoxu Wang, Wenxiang Dong, Wenyi Zhang, and Chee~Wei Tan.
\newblock Rumor source detection with multiple observations: Fundamental limits
  and algorithms.
\newblock In {\em ACM SIGMETRICS Performance Evaluation Review}, volume~42,
  pages 1--13. ACM, 2014.

\end{thebibliography}
	
	\pagebreak
	\nobalance
	\appendix
\section{Complementary results}

\subsection{Properties of the binary tree}
We first establish a few properties of the complete binary tree.

\begin{proposition} \label{cl:Wtree} 
The {\sc CutWidth} of the complete binary tree is smaller than $\log(N)$.
\proof Consider the crusade \cite{Drakopoulos2015} implied by a Deep First Search over the tree. This crusade has a maximal cut of $\log(N) - 1$. Thus, by definition, the {\sc CutWidth} is lower than $\log(N) - 1$.

\end{proposition}

\begin{proposition} \label{cl:Nsubtrees} 
There are $r^3$ subtrees containing $\frac{N}{r^3}$ nodes, and they are at distance $\OO(\log\log(N))$ from the root.
\proof In the complete binary tree, there are $2^k$ subtrees at distance $k$ from the root that contain $\frac{N}{2^k}$ nodes. The results follows for $k = \frac{3\log(r)}{\log(2)}$. \\\qed 
\end{proposition}

\subsection{Some probabilities}

\subsubsection{Geometric variables}
\begin{proposition} \label{cl:minGeo} 
The minimum of $i$ independent geometric random variables of parameter $\mu$ is a geometric random variable of parameter $1 - (1-\mu)^i$.
\proof Let $Geo(i,\mu)$ be the minimum of $i$ independent geometric random variables. Then:
\begin{align*}
\p(Geo(i,\mu) \geq k) &= ((1 - \mu)^{k-1})^i \\
&= ((1 - \mu)^i)^{k-1}
\end{align*}
We recognize the probability distribution of a geometric variable with parameter $1 - (1-\mu)^i$.
\end{proposition}

\begin{lemma} \label{lem:infectionTime} 
As $\tau \to 0$, it takes less than $\frac{\log(k)}{\tau} $ time steps in expectation to infect $k$ new nodes.
\proof Every new infection increases the cut by 1. Let $Geo(i, \mu)$ be the minimum of $i$ geometric random variables of parameter $\mu$, and let $T_k$ be the time it takes to infect the $k$ new nodes. We have:
\begin{align*}
T_k &= \sum_{i=1}^{k-1} Geo(i, \mu)
\end{align*}
Using Claim \ref{cl:minGeo}, $Geo(i,\mu)$ is a geometric variable of parameter $1- (1-\mu)^i$. Therefore:
\begin{align*}
\E(T_k) &= \E\left(\sum_{i=1}^{k-1} Geo(i, \mu)\right) \\
&= \sum_{i=1}^{k-1} \frac{1}{1-(1-\mu)^i} \\
&=_{\tau \to 0} \sum_{i=1}^{k-1} \frac{1}{i\tau} \\ 
&\leq_{\tau \to 0} \frac{\log(k)}{\tau}
\end{align*}
\\\qed 
\end{lemma}

\subsubsection{Some curing probabilities}

\begin{proposition} \label{cl:curingDelta} 
	If all the budget at a given time step is spent, the probability that no nodes are cured in this time step is $1-\delta$.
	\proof Let $r_i$ be the budget attributed to the node $i$. Then:
	\begin{align*}
	P_{\rm NoCuring} &= \prod_{i=1}^{N} 1 - \delta_i = \prod_{i=1}^{N} e^{-r_i \tau} \\
	&= e^{-\left(\displaystyle\sum_{i=1}^{N} r_i \right) \tau} = e^{-r \tau} \\
	&= 1 - \delta.
	\end{align*}
	\qed 
\end{proposition}

\begin{lemma}\label{lem:Ppath} 
	The probability $P^{\rm path}_{\rm m}$ that $m$ nodes are reinfected along a path, such that no node on the $m$-length path is cured before they all become infected, is lower bounded by $\left(\frac{\mu (1-\delta)}{\delta + \mu (1-\delta)} \right)^{m+1}$.
	\proof Using Proposition \ref{cl:curingDelta}, 
	
	\begin{align*}
	P^{\rm path}_{\rm m} &\geq \sum_{t=0}^{\infty}  {m + t \choose m}  \mu^{m+1} (1-\mu)^t (1-\delta)^{m + t} \\
	&\geq (\mu (1-\delta))^{\rm m} \cdot \mu \cdot \sum_{t=0}^{\infty}  {m + t \choose m}   \left((1-\mu)(1-\delta)\right)^t \\
	&\geq (\mu (1-\delta))^{m} \cdot \mu \cdot \frac{1}{\left(1- (1-\mu)(1-\delta))\right)^{m+1}} \\
	&\geq \left(\frac{\mu (1-\delta)}{\delta + \mu (1-\delta)} \right)^{m+1}.
	\end{align*}
	\qed 
\end{lemma}

\begin{corollary} \label{cor:Ppath} 
	The probability $P^{\rm startPath}_{\rm m}$ that $m$ nodes are reinfected along a path, such that no node on the $m$-length path is cured before they all become infected, and such that there is an infection on the first time step, is lower bounded by $\mu\cdot \left(\frac{\mu (1-\delta)}{\delta + \mu (1-\delta)} \right)^{m}$.
	\proof Taking into account that the first time step is an infection:
	\begin{align*}
	P^{\rm startPath}_{\rm m} &\geq \sum_{t=0}^{\infty}  {m - 1 + t \choose m - 1}  \mu^{m+1} (1-\mu)^t (1-\delta)^{m + t} \\
	&\geq \mu \cdot\left(\frac{\mu (1-\delta)}{\delta + \mu (1-\delta)} \right)^{m}.
	\end{align*}
	\qed 
\end{corollary}

\begin{proposition}\label{cl:chernoffEndgame} 
	Let $T_{\frac{N}{2r^4}}$ be the random variable representing the time to cure half of the $\frac{N}{r^4}$ last nodes. Then: 
	$$\p\left(T_{\frac{N}{2r^4}} \leq \frac{N}{4r^5\delta}\right) \leq e^{-\frac{N}{8r^5}}.$$
	\proof The difficulty here lies in the fact that we want to obtain exponential concentration inequalities on a sum of geometric variables, which are unbounded. Therefore, we cannot directly use a Chernoff's bound. Following an idea from ~\cite{Brown}, we represent geometric variables as the sum of Bernoulli variables. Each variable is then bounded, which makes the analysis possible.
	
	Let $X_i^t$ be 1 if node $i$ was cured at time $t$, and 0 otherwise. Let $X^t$ be $r$ with probability $\delta$, and 0 otherwise. We notice $\p(X_i^t = 1) \leq \delta$, and $\forall t, \sum_{i=1}^{N} X_i^t \leq r$. By using Chernoff's bound on a sum of $\frac{N}{4r^5\delta}$ Bernoulli variables of parameter $\delta$, we can therefore bound the probability that curing $\frac{N}{2r^4}$ nodes happens in a short time (here less than $\frac{N}{4r^5\delta}$ time):
	\begin{align*}
	\p(T_{\frac{N}{2r^4}} \leq \frac{N}{4r^5\delta}) &= \p(\sum_{t=1}^{\frac{N}{4r^5\delta}} \sum_{i=1}^{N} X_i^t \geq \frac{N}{2r^4}) \\
	&\leq \p(\sum_{t=1}^{ \frac{N}{4r^5\delta}} X^t \geq \frac{N}{2r^4}) \\
	&\leq \p(\sum_{t=1}^{ \frac{N}{4r^5\delta}} \frac{X^t}{r} \geq \frac{N}{2r^5}) \\
	&\leq \p\left(\sum_{t=1}^{ \frac{N}{4r^5\delta}} \frac{X^t}{r} \geq \frac{N}{4r^5\delta}\cdot \delta \cdot (1+1)\right) \\
	&\leq \p\left(\sum_{t=1}^{ \frac{N}{4r^5\delta}} \frac{X^t}{r} \geq \E\left[\sum_{t=1}^{ \frac{N}{4r^5\delta}} \frac{X^t}{r} \right]\cdot (1+1)\right) \\
	&\leq e^{-\frac{ \frac{N}{4r^5}\cdot 1^2}{3}} \\
	&\leq e^{-\frac{N}{12r^5}}.
	\end{align*}
	\qed
\end{proposition}

\begin{proposition} \label{cl:noOtherInfection} 
	Conditioned on reaching a cut of $3r$ in a minimal tree in less than $\frac{30\log(r)}{\tau}$ time steps, the probability of not infecting any nodes outside of the $escape$ $P_{\rm NoOtherInfections}$ is bounded by:
	\begin{align*}
	P_{\rm NoOtherInfections} &\leq e^{-\frac{360\log^2(r)\mu}{\tau}} \\
	&\leq_{\tau \to 0} e^{-360\log^2(r)}.
	\end{align*}
	\proof
	For an infection to not be part of the $Escape$, it has to happen because of a node which is either on the path to the root, or on the path to a minimal tree. As calculated before, there are $12\log(r)$ such nodes, which all have at most one edge not on the path (the two others were used either to get infected, or to infect the next node on the path). What's more, each of these nodes was infected for at most $\frac{30\log(r)}{\tau}$ time steps. The probability of not infecting any node along those edges during all these time steps is therefore:
	\begin{align*}
	P_{\rm NoOtherInfections} &\leq \left((1-\mu)^{12\log(r)} \right)^{\frac{30\log(r)}{\tau}} \\
	&\leq (1-\mu)^{\frac{360\log^2(r)}{\tau}} \\
	&\leq e^{-\frac{360\log^2(r)\mu}{\tau}}.
	\end{align*}
	As $\tau$ goes to 0:
	\begin{align*}
	P_{\rm NoOtherInfections} &\leq  e^{-\frac{360\log^2(r)\mu}{\tau}} \\
	&\leq_{\tau \to 0} e^{-360\log^2(r)}.
	\end{align*}
	\qed
\end{proposition}

\subsubsection{Moment generating function of the random walk}

\begin{proposition} \label{cl:MGF}
There exists $x^*>0$ such that the Moment Generating Function (MGF) of $G_t$ evaluated at $x^*$ is 1.
\proof Since $G_t$ is a sum of independent random variables, and since the MGF of a Bernouilli random variable of parameter $p$ is equal to $MGF(x) = pe^x +(1-p)$:
\begin{align*}
MGF_{G_t}(x) &= MGF_{C_t}(x) \cdot MGF_{I_t}(x) \\
&= (\delta e^x + (1 - \delta))^r\cdot (\mu e^{-x} + (1 - \mu))^\frac{r^3}{3}
\end{align*}
We can see that:
 $$MGF(0) = 1, \qquad MGF'(0) < 0, \qquad MGF(r) \to_{r \to \infty} \infty$$
Therefore, by the Intermediate Value Theorem:
$$ \exists x^* > 0, MGF(x^*) = 0 $$
There is no closed form solution for $x^*$, but we can get an approximation when $\tau \to 0$. \\\qed 
\end{proposition}

\begin{proposition} \label{cl:MGFtau0} 
When $\tau \to 0$, we have a closed form solution: $x^* = \log(r) - \log(3)$.
\proof When $\tau \to 0$:

\begin{align*}
1 &= MGF_{G_t}(x^*) = (\delta e^{x^*} + (1 - \delta))^r\cdot (\mu e^{-x^*} + (1 - \mu))^\frac{r^3}{3} \\
&=_{\tau \to 0} (r\tau e^{x^*} + (1 - r\tau))^r\cdot (\tau e^{-x^*} + (1 - \tau))^\frac{r^3}{3} \\
&=_{\tau \to 0} (1 + r\tau (e^{x^*} - 1))^r (1 + \tau(e^{-x^*} - 1))^\frac{r^3}{3} \\
&=_{\tau \to 0} (1 + r^2\tau (e^{x^*} - 1)) (1 + \tau \frac{r^3}{3} (e^{-x^*} - 1)) \\
&=_{\tau \to 0} 1 + \tau \left( r^2(e^{x^*} - 1) - \frac{r^3}{3} (1 - e^{-x^*}) \right) \\
&=_{\tau \to 0} 1 + \tau \left( r^2 \cdot e^{-x^*}(e^{2x^*} - e^{x^*} - \frac{r\cdot e^{x^*}}{3}  + \frac{r}{3}) \right)
\end{align*}
If we want to nullify the first order in $\tau$, we need:

$$  (e^{x^*})^2  - (1 + \frac{r}{3}) (e^{x^*}) + \frac{r}{3} = 0 $$
This is a second order polynomial, which gives us the solution $e^{x^*} = 1$ (trivial solution for $x^* = 0$), and $e^{x^*} = \frac{r}{3}$, which gives us a non-trivial solution:

$$x^* = \log(r) - \log(3) > 0 $$ \\\qed 
\end{proposition}

\subsection{Some calculus}
\subsubsection{monotonicity results}

\begin{proposition} \label{cl:taudelta} 
The function $k(x) = \frac{x}{1-e^{-rx}}$ is increasing in x. In particular, for all $x \leq 0$, we have $k(x) \geq k(0) = \frac{1}{r}$.
\proof $x \to x$ and $x \to \frac{1}{1-e^{-rx}}$ are both increasing functions of $x$, so $k(x)$ is also increasing. \\\qed 
\end{proposition}

\section{A policy achieving the upper bound} \label{sec:upperBound}
The main contribution of this paper proves a lower bound on the budget in the Partial information setting. We prove that for budget $r = \OO(\log(N))$, there exists no strategy which allows polynomial expected curing time, unless $D(p||q)/\tau$ goes to infinity. Moreover, our result implies that if $D(p||q)/\tau = 0$, then for budget $r = \OO({\rm poly}(\log(N)))$, there exists no strategy which allows polynomial expected curing time. \\\\
We now study the converse problem. In this section we exhibit a policy which:
\begin{itemize}
	\item Does not require any knowledge of the state (works even in the Blind Curing setting);
	\item Achieves linear expected curing time;
	\item Needs $r \sim \OO(e^{\frac{4}{c}} \cdot N^c)$ budget, for any $c>0$.
\end{itemize}

\subsection{Description of the policy}
We consider the ordering $O$ of the nodes given by a Depth First Search on a binary tree. We split the graph into 3 sets: $A_{{\rm sus}}$, $A_{{\rm inf}}$ and $A_{{\rm buff}}$. Intuitively, these sets respectively represent the set of the nodes we believe are cured, the set of nodes we believe are infected, and the buffer zone in the middle. 

We run through the following algorithm. As we show in Section \ref{ssec:combining_upperbd}, the probability that we fail to cure the graph in one pass of the algorithm below (what we call one {\em iteration}), is at most $2/N$, and hence the expected time to cure, given our budget, is linear. 

To initialize each pass of the algorithm, we set $t=0$, and also initialize the sets $A_{{\rm sus}}^0 = A_{{\rm buff}}^0 = \emptyset$, $A_{{\rm inf}}^0 = V$.\\
Every $\frac{1}{\tau}$ time steps, we:
\begin{itemize}
	\item move a node from $A_{{\rm inf}}$ to $A_{{\rm buff}}$, following the ordering $O$
	\item remove all the nodes from $A_{{\rm buff}}$ which are at distance greater than $c\log(N)$ from any node of $A_{{\rm inf}}$, and place them in $A_{{\rm sus}}$
\end{itemize}
Then, during $\frac{1}{\tau}$ time steps, we:
\begin{itemize}
	\item cure all the nodes of $A_{{\rm buff}}$ with constant budget $c_1$;
	\item cure the new node with budget $(1 + c_2) \log(N)$, where $c_2$ constant.
\end{itemize}

This gives a total budget of $2^{c\log(N)}\cdot c_1 + c_2\log(N) = N^{c\log(2)} \cdot c_1 + c_2 \log(N)$.

At time step $t=\frac{N}{\tau}$, when $A_{{\rm inf}}^{\frac{N}{\tau}} = \emptyset$, we keep curing $A_{{\rm buff}}^{\frac{N}{\tau}}$ for an additional $\frac{c\log(N)}{\tau}$ time steps. 

One pass through the set of actions described above is called an \textbf{iteration}. We show below that the probability of failing to cure the entire graph in one iteration is bounded by $2/N$, and hence we can cure the graph in linear expected time. Equivalently, in time $\gamma \cdot N$, one can get a $(1-\epsilon)$-probability guarantee that the graph is cured, for any $\epsilon > 0$. 


\subsection{Properties of the policy}
\begin{proposition}{Every node of the graph spends at least $\frac{c\log(N)}{\tau}$ time steps in $A_{{\rm buff}}$.}
\proof We notice $A_{{\rm inf}}$ is connected at all time. Therefore, when a node $i$ is removed from $A_{{\rm inf}}$ and added to $A_{{\rm buff}}$, it is at distance 1 from a node of $A_{{\rm inf}}$. Every subsequent node transferred from $A_{{\rm inf}}$ to $A_{{\rm buff}}$ can only increase the distance between $i$ and $A_{{\rm inf}}$ by 1. Since a new node is transferred every $\frac{1}{\tau}$ time steps, and all the nodes at distance no greater than $c\log(N)$ from $A_{{\rm inf}}$ are kept in $A_{{\rm buff}}$, every node $i$ of the graph spends at least $\frac{c\log(N)}{\tau}$ time steps in $A_{{\rm buff}}$.
\\\qed 
\end{proposition}

\begin{proposition}{Let $T_{{\rm cured}}$ be the time it takes to cure the graph, and $P_{{\rm OneIteration}}$ be an lower bound on the probability that the graph is cured in one iteration. Then:
	$$\E[T_{{\rm cured}}] \leq \frac{N + c\log(N)}{P_{{\rm OneIteration}}}.$$}
\proof $T_{{\rm cured}}$ is stochastically dominated by an exponential variable with parameter $P_{{\rm OneIteration}}$, which in turn has expectation $\frac{1}{P_{{\rm OneIteration}}}$. One iteration lasts exactly $\frac{N}{\tau} + \frac{c\log(N)}{\tau}$ time steps, and one time step lasts $\tau$ time, so an iteration lasts $N + c\log(N)$ time.
\\\qed 
\end{proposition}

\subsection{Analysis}
\begin{definition}
	We call an \textbf{epoch} $\frac{1}{\tau}$ consecutive time steps.
\end{definition}

If there is at least one infected node at the end of the policy, then either one of the following events must have happen:
\begin{enumerate}
	\item One node was not cured when it entered the buffer zone, and then proceeds to make its way to $A_{{\rm sus}}$.
	\item There was a path of infection from a node of $A_{{\rm inf}}$ to a node of $A_{{\rm sus}}$.
\end{enumerate}
We calculate the probability of the two events above happening during one epoch:

\subsubsection{Case 1: One node was not cured when it entered the buffer zone, and then proceeds to make its way to $A_{{\rm sus}}$.}
The probability of this event is lower than the probability that one node was not cured during one epoch when it entered the buffer zone:
\begin{align*}
\p(\rm{Case 1}) &\leq (1 - (1- e^{-(1+c_2)\log(N) \tau}))^{\frac{1}{\tau}} \\
&\leq e^{-(1+c_2)\log(N)} \\
&\leq \frac{1}{N^{1+c_2}}.
\end{align*}

\subsubsection{Case 2: There was a path of infection from a node of $A_{{\rm inf}}$ to a node of $A_{{\rm sus}}$.}
In the case 2:
\begin{enumerate}
	\item One node $n_s$ of $A_{buf}^{t_0}$ needs to become infected at time step $t_0$.
	\item One node $n_e$ of $A_{{\rm sus}}^{t_0 + t}$ becomes infected after $t$ time steps.
	\item Every $\frac{1}{\tau}$ time steps, the nodes of $A_{{\rm sus}}$ can become closer to where the infected node by a distance $1$. Therefore, $c\log(N) -\lfloor \tau\cdot t \rfloor - 1$ additional infections need to happen along the unique path between $n_s$ and $n_e$.
\end{enumerate}
Let us calculate the probability $p_1$ that $b$ becomes infected at time $t_0$, and then proceed to infect $c\log(N) -\lfloor \tau\cdot t \rfloor$ additional nodes along a cured path in  $t$ time steps, with the head of the infection not being cured:
\begin{align*}
p_1 = \mu \cdot {t \choose \max(0, c\log(N) -\lfloor \tau\cdot t \rfloor  - 1)} \mu^{c\log(N) -\lfloor \tau\cdot t \rfloor}(1-\beta)^{t}.
\end{align*}
Let us now sum over all time steps $t$, to get the probability $p_2$ that an infection reaches $A_{{\rm sus}}$ with exactly $d$ infections, starting from one time step:

\begin{align*}
p_2 &= \mu \cdot \displaystyle\sum_{t=\frac{c\log(N)}{1+\tau}}^{\infty}  {t \choose \max(0, c\log(N) -\lfloor \tau\cdot t \rfloor - 1)}  \mu^{c\log(N) -\lfloor \tau\cdot t \rfloor}(1-\beta)^{t} \\
&\leq \mu \cdot \displaystyle\sum_{t=\frac{c\log(N)}{1+\tau}}^{\infty}   {t \choose c\log(N) - 1}\frac{(c\log(N))!}{(c\log(N) -\lfloor \tau\cdot t \rfloor)!}\frac{t!}{(t+ \tau t)!} \\
&\qquad \cdot \mu^{c\log(N) -\lfloor \tau\cdot t \rfloor}(1-\beta)^{t} \\
&\leq \mu \cdot \displaystyle\sum_{t=\frac{c\log(N)}{1+\tau}}^{\infty}  {t \choose c\log(N) - 1}\left(\frac{c\log(N)}{t+ \lfloor \tau t \rfloor}\right)^{\lfloor \tau t\rfloor } \mu^{c\log(N) -\lfloor \tau\cdot t \rfloor}(1-\beta)^{t} \\
&\leq \mu \cdot \displaystyle\sum_{t=c\log(N)}^{\infty}  {t \choose c\log(N) - 1} \mu^{c\log(N) - \tau\cdot t}(1-\beta)^{t} \\
&= \mu \cdot \displaystyle\sum_{t'=0}^{\infty} {c\log(N)-1+t' \choose c\log(N)-1} \mu^{c\log(N)}\left(\frac{1-\beta}{\mu^\tau}\right)^{c\log(N) - 1 + t'} \\
&= \mu^{c\log(N) + 1}  \cdot \left(\frac{1-\beta}{\mu^\tau}\right)^{c\log(N) - 1} \frac{1}{\left(1-\left(\frac{1-\beta}{\mu^\tau}\right)\right)^{c\log(N)+1}} \\
&\sim_{\tau \to 0} \tau^{c\log(N) + 1}  \cdot 1^{c\log(N) - 1}  \frac{1}{\left(1-\left(\frac{1-c_1\cdot \tau}{\tau^{\tau}}\right)\right)^{c\log(N)}} + o(\tau)\\
&\sim_{\tau \to 0} \frac{\tau^{c\log(N) + 1}}{(c_1 \tau)^{c\log(N)}} + o(\tau) \\
&\sim_{\tau \to 0} \frac{\tau}{c_1^{c\log(N)}} + o(\tau)\\
&\sim_{\tau \to 0} \frac{\tau}{N^{c\cdot\log(c_1)}} + o(\tau),
\end{align*}
where we have used that $\left(\frac{c\log(N)}{t+ \lfloor \tau t \rfloor}\right) < 1$, that $\mu < 1$, so $\mu^{-\lfloor \tau\cdot t \rfloor} \leq \mu^{- \tau\cdot t}$, that $\displaystyle\sum_{k=0}^{\infty} {m + k \choose k} a^{k} = \frac{1}{(1-a)^{m+1}}$ when $|a| < 1$, and that $\tau^\tau \to_{\tau \to 0} 1$.

Now, if we select a starting node $n_s$ and an end node $n_e$, there is only one path between them in a tree. Such an infection can start $\frac{1}{\tau}$ times during one epoch. We can therefore apply a union bound:

\begin{align*}
\p(\rm Case 2) &\leq \sum_{t_0 = 1}^{\frac{1}{\tau}} \sum_{n_s, n_e} p_2 \\
&\leq \frac{N^2}{\tau} \cdot p_2 \\
&\leq \frac{1}{N^{c\cdot\log(c_1) - 2}}.
\end{align*}

\subsection{Combining the results for all time steps}
\label{ssec:combining_upperbd}
At each epoch, the probability of failure is upper bounded by $\p(\rm Case 1) + \p(\rm Case 2)$. The probability of failing during one iteration, which lasts $N + c\log(N)$ epochs, is therefore:
\begin{align*}
\p(\rm {\rm OneIteration}Fail) &\leq  (N + c\log(N))\cdot (\p(\rm Case 1)  + \p(\rm Case 2)) \\
&\leq 2N \cdot (\frac{1}{N^{c\cdot\log(c_1) - 2}} + \frac{1}{N^{1+c_2}}).
\end{align*}

Therefore, if we choose $c_2 = 1$, and $c_1 = e^{\frac{4}{c}}$, we have:

\begin{align*}
\p(\rm {\rm OneIteration}Fail) &\leq 2N \cdot (\frac{1}{N^{4 - 2}} + \frac{1}{N^{1+1}}) \\
&\leq  \frac{2}{N}.
\end{align*}

We have, therefore, an upper bound, as stated in Theorem \ref{thm:new_upperbound}. We repeat here, and complete the proof.
\vspace{0.2cm}

\noindent
{\sc Theorem} \ref{thm:new_upperbound}. {\em
	In the Blind Curing setting, for all $c>0$, we can cure the binary tree in expected linear time with budget $\OO(e^{\frac{4}{c}}\cdot N^{c} )$.
	\proof If we choose $c_2 = 1$, and $c_1 = e^{\frac{4}{c}}$, we have:
	
	\begin{align*}
	\p(\rm {\rm OneIteration}Fail) &\leq 2N \cdot (\frac{1}{N^{4 - 2}} + \frac{1}{N^{1+1}}) \\
	&\leq  \frac{2}{N}.
	\end{align*}
	
	Therefore, with budget $e^{\frac{4}{c}}\cdot N^{c} + \log(N)$, for all $c>0$:
	\begin{align*}
	\E[T_{{\rm cured}}] &\leq \frac{N + c\log(N)}{1 - \frac{2}{N}} \\
	&\leq 4N.
	\end{align*}
	}

\section{Numerical experiments} \label{sec:experiments}
In this section, we add some numerical experiments to illustrate the difficulty of the problem. We introduce two curing strategies: Naive Curing, which cures randomly a subset of nodes which signal themselves as infected (they raise a flag), and the strategy from Section \ref{sec:upperBound}, Blind Protection, which prevents the infection from spreading by curing every node near the infected set. We hope these two strategies can provide insight into the difficulty of curing the complete binary tree in our model. 

It is important to understand that the results we present are strategy-specific, which means better results could possibly be achieved with better strategies. Devising optimal strategies is however outside of the scope of this work.

\subsection{Impact of the lack of information}
In this section, we illustrate the dramatic impact of the lack of information on the Naive Curing strategy. In the following experiment, we consider a binary tree on 31 nodes. We use a budget $r = 16>\frac{N}{2}$. If $p_\epsilon$ is the probability of error, at each time step, an infected node raises a flag with probability $1 - p_\epsilon$, and a susceptible node raises a flag with probability $p_\epsilon$. We set the size of a time step to be $\tau = 0.1$.

\begin{figure}[H]
	\centering\includegraphics[width=8cm]{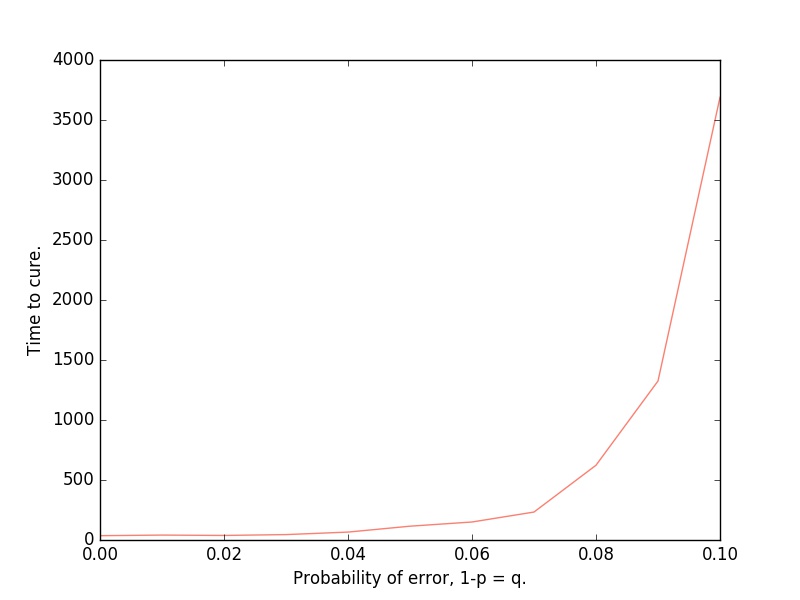}
	\caption{Time to cure as a function of the probability of error for the Naive Curing strategy.}
	\label{fig:naive}
\end{figure}

The results can be seen in Figure \ref{fig:naive}. The time to cure increases faster than exponentially with the probability of error. We can see that even with 10\% of error, it takes more than 3500 time steps with budget $r= \frac{N}{2}$ on 31 nodes.

\subsection{Impact of size of the graph}
We now consider the strategy described in Appendix \ref{sec:upperBound}. For the purpose of these experiments, keeping the same notation as the previous section, we set $c_1 = 10$ (this is the budget for every node which we "protect"), and $c_2 = 1$ (we cure any new node with budget $(1+c_2)\log(N)$). We still have $\tau = 0.1$. For this experiment, we investigate the time it takes to cure the graph for a budget equal to different exponents of the number of nodes.

\begin{figure}[H]
	\centering\includegraphics[width=8cm]{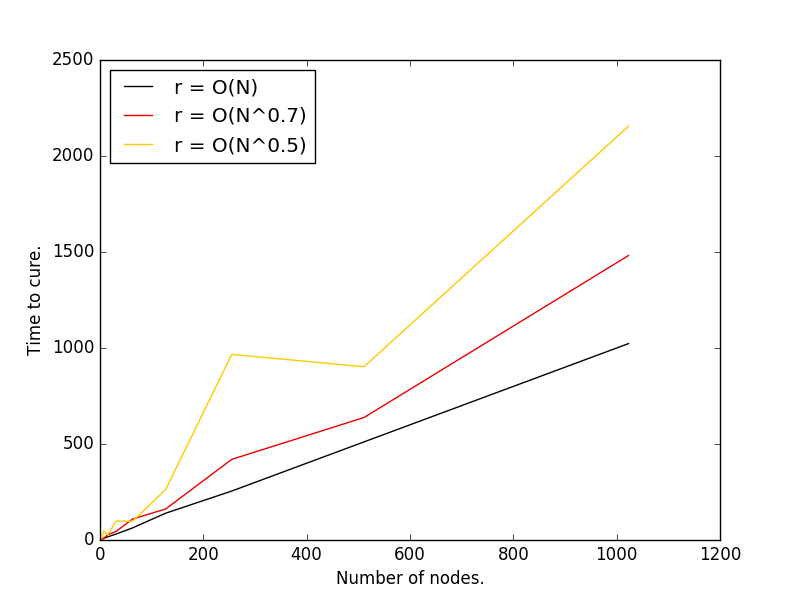}
	\caption{Time to cure as a function of the number of nodes for the Blind Protection strategy. The plots are the average of 20 runs.}
	\label{fig:Npower}
\end{figure}

The results are shown in Figure \ref{fig:Npower}. As theory predicts, the time to cure increases more slowly than $4\cdot N$, where $N$ is the number of nodes for budget $r = \OO(N^c)$, for all $c>0$ constant. 

As a reminder, in the Blind Curing setting, it is impossible to cure the complete binary tree in less than superpolynomial time for budget $r=\OO(\log^\alpha(N))$, for all $\alpha > 0$ constant.

\end{document}